\documentclass[10pt, twocolumn, english, amssymb, aps, pra, superscriptaddress, showpacs, showkeys, floatfix]{revtex4-2}

\usepackage{booktabs}
\usepackage{amsmath, amssymb, graphicx}
\usepackage{mathptmx}
\usepackage{physics}
\usepackage{xcolor}
\usepackage{textcomp}
\usepackage{lineno}
\usepackage[utf8]{inputenc}
\usepackage[scr=rsfso, cal=zapfc, frak=euler, bb=ams]{mathalfa}
\usepackage{enumitem}
\usepackage{tikz}
\usetikzlibrary{arrows.meta, positioning, calc}
\usepackage{tabularx}

\usepackage{array,tabularx,ragged2e}
\newcolumntype{Y}{>{\RaggedRight\arraybackslash}X}

\usepackage{hyperref}

\begin{document}

\title{
Collective-Coordinate Fluctuations of Driven-Dissipative Solitons}

\author{Yifan Sun}
\email{yifan.sun@ulb.be}
\affiliation{Service OPERA-Photonics, Université libre de Bruxelles, B-1050 Brussels, Belgium}
\author{Thomas Bunel}
\affiliation{Service OPERA-Photonics, Université libre de Bruxelles, B-1050 Brussels, Belgium}
\author{Sofya Glazyrina}
\affiliation{Service OPERA-Photonics, Université libre de Bruxelles, B-1050 Brussels, Belgium}
\author{Georges Semaan}
\affiliation{Service OPERA-Photonics, Université libre de Bruxelles, B-1050 Brussels, Belgium}
\author{Fabien Bretenaker}
\affiliation{Université Paris-Saclay, ENS Paris-Saclay, CNRS, CentraleSupélec, LuMIn, Orsay, France}
\author{Stephane Coen}
\affiliation{Physics Department, The University of Auckland, Auckland 1010, New Zealand}
\affiliation{The Dodd-Walls Centre for Photonic and Quantum Technologies, Auckland 1010, New Zealand}
\author{Simon-Pierre Gorza}
\affiliation{Service OPERA-Photonics, Université libre de Bruxelles, B-1050 Brussels, Belgium}
\author{François Leo}
\affiliation{Service OPERA-Photonics, Université libre de Bruxelles, B-1050 Brussels, Belgium}

\begin{abstract}
Fluctuations of nonequilibrium localized waves are shaped not only by direct stochastic forcing but also by deterministic transfer among coupled collective degrees of freedom. We develop a pathway-resolved stochastic collective-coordinate theory that makes this transfer explicit for stationary driven-dissipative solitons of the generalized Lugiato--Lefever equation with Raman response. 
The reduction yields a refined stationary phase-locking relation, providing a fixed point for the subsequent stochastic theory.
Projecting field-level fluctuations onto four soliton coordinates: amplitude, frequency shift, temporal position, and global phase, yields a reduced Langevin model and, after linearization about a stable stationary state, an analytic power-spectral-density matrix. 
This framework separates direct stochastic injection from deterministic inter-coordinate conversion and thereby resolves how each observable spectrum is assembled from distinct internal fluctuation pathways. It shows that timing jitter is governed primarily by Gordon--Haus-type frequency-to-timing conversion, while phase noise is often dominated by amplitude-to-phase transfer rather than by direct phase diffusion. Raman response opens additional cascaded pathways, and the low-detuning hump in the intensity and phase spectra is traced to the driven response of an underdamped amplitude--phase subsystem preceding the breathing instability. Comparisons with stochastic simulations of both the reduced model and the full generalized Lugiato--Lefever equation show good agreement throughout most of the stable stationary single-soliton regime, with systematic deviations mainly near the Hopf boundary. The theory provides a general route for connecting internal fluctuation-transfer mechanisms of dissipative solitons to measurable noise observables.
\end{abstract}

\maketitle

\section{Introduction}
\label{sec:introduction}

In nonequilibrium nonlinear systems, fluctuations around stable localized states are shaped not only by stochastic forcing, but also by the deterministic linearized dynamics around the underlying attractor \cite{crossPatternFormationOutside1993,greluDissipativeSolitonsModelocked2012}. 
For optical solitons, this idea appears particularly clearly in collective-coordinate descriptions, where perturbations of amplitude, frequency, timing, and phase are dynamically coupled \cite{gordonRandomWalkCoherently1986,hasegawaOpticalSolitonsFibers2003,Haus1993,Paschotta2004}. 
As a result, the noise observed in a measured output need not reflect the direct forcing of that observable alone; fluctuations can be transferred, filtered, and redistributed internally before appearing in experimentally accessible quantities. 
For dissipative coherent structures, the central question is therefore not only how much noise is present, but through which internal pathways it reaches a given observable.

This viewpoint already has familiar precedents in nonlinear optics. In conservative fiber solitons, Gordon--Haus jitter is the canonical example of indirect fluctuation transfer: noise-induced frequency perturbations are converted into arrival-time fluctuations by the soliton dynamics \cite{gordonRandomWalkCoherently1986,hasegawaOpticalSolitonsFibers2003}. Closely related couplings among amplitude, frequency, timing, and optical phase have long been recognized in mode-locked lasers \cite{Haus1993,Paschotta2004,Kim2016reviewModelockedLasers,Fortier2019_20YearsDevelopments}. 
Phase--amplitude noise correlations have also been resolved experimentally in optical frequency combs \cite{ansquerUnveilingDynamicsOptical2021}.
These couplings underlie carrier-envelope-offset and comb-line phase-noise dynamics \cite{Helbing2002a,Paschotta2006}, as well as more platform-specific noise analyses in semiconductor, fiber, and nanolaser systems \cite{Jiang2001,Paschotta2010,Kim2016reviewModelockedLasers,sunNoiseInvestigationCW2022}. 
What remains less well developed for driven-dissipative solitons is a framework that resolves how the spectrum of a measured observable is assembled from direct stochastic injection and inter-coordinate transfer.

Temporal cavity solitons provide a particularly stringent setting for this problem. They are localized attractors of a coherently driven, lossy nonlinear resonator and support both a small set of physically meaningful collective coordinates and directly measurable optical and microwave observables. At the mean-field level, their dynamics are governed by the driven--damped nonlinear Schr\"odinger equation, commonly written in this context as the temporal Lugiato--Lefever equation (LLE) \cite{lugiatoSpatialDissipativeStructures1987,wabnitzSuppressionInteractionsPhaselocked1993,coenModelingOctavespanningKerr2013,Coen2013Scaling}. The LLE supports stationary localized states, breathers, and destabilized solutions \cite{godeyStabilityAnalysisSpatiotemporal2014,sunDynamicsDissipativeStructures2023}, and temporal cavity solitons have been observed in both fiber-loop cavities and integrated microresonators \cite{leoTemporalCavitySolitons2010,herrTemporalSolitonsOptical2014}. 
Because they now form the dynamical core of the soliton-based frequency comb platform \cite{kippenbergDissipativeKerrSolitons2018}, including low-noise architectures for optical frequency division and microwave synthesis \cite{DelHaye2016PhaseCoherent,lucasUltralownoisePhotonicMicrowave2020}, synchronization \cite{moilleKerrinducedSynchronizationCavity2023}, and phase-stabilized operation \cite{wildiPhasestabilisedSelfinjectionlockedMicrocomb2024}, understanding how internal soliton fluctuations are converted into intensity, frequency, timing, and phase noise is both a fundamental and practical problem.

\begin{figure*}[htb!]
    \centering
    \includegraphics[width=1\textwidth]{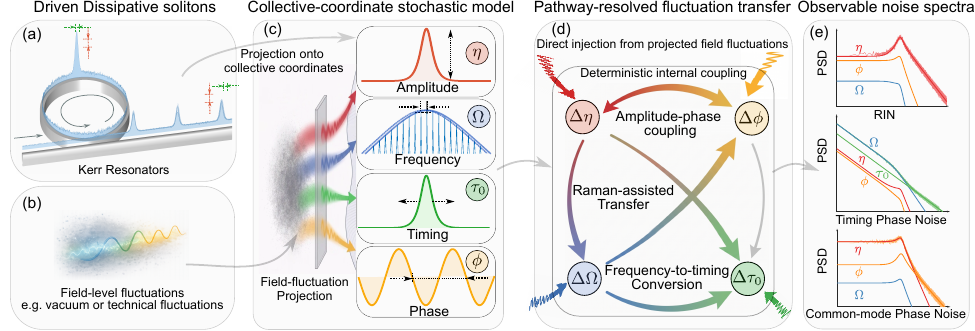}
    \caption{
Concept of the pathway-resolved stochastic framework for stationary driven-dissipative cavity solitons. 
Complex field (a) and Field-level fluctuations (b) are projected onto a reduced soliton manifold spanned by four collective coordinates (c): amplitude $\eta$, frequency shift $\Omega$, temporal position $\tau_0$, and global phase $\phi$. 
These projected fluctuations then enter the reduced model as direct stochastic injection into the collective coordinates, after which deterministic internal coupling (d) redistributes them through pathway-resolved transfer channels, including amplitude-phase coupling, frequency-to-timing conversion, and Raman-assisted transfer. 
The resulting collective-coordinate fluctuations are finally mapped onto observable noise spectra (e), including relative intensity noise (RIN), timing phase noise, and common mode phase noise.
}
    \label{fig:toy_overview}
\end{figure*}

To formulate such a pathway-resolved fluctuation theory, the minimal LLE is a useful starting point, but it is not always the most representative setting for experimentally relevant cavity solitons. 
Additional deterministic perturbations can modify both the stationary soliton state and the internal couplings through which fluctuations are routed between observables. 
For example, higher-order dispersion and dispersive-wave recoil can break spectral symmetry, reorganize existence and stability, and underlie noise-conversion effects in soliton microcombs \cite{parra-rivasThirdorderChromaticDispersion2014,braschPhotonicChipbasedOptical2016,liExperimentalObservationsBright2020,yangDispersivewaveInducedNoise2021,leiOpticalLinewidthSoliton2022}. 
Structured driving, trapping potentials, and intracavity phase landscapes provide another route to controlling or stabilizing cavity solitons, thereby modifying the sensitivity of timing, repetition rate, and related observables to fluctuations \cite{Erkintalo2022,sunDissipativeKerrSolitons2022,sunMultimodeResonanceTransition2023,talentiControlStabilizationKerr2023,englebertDynamicsDrivenDissipative2026a}. 
Here we focus on Raman response as a minimal nontrivial extension of the LLE. 
Raman response shifts the soliton carrier, couples amplitude and frequency dynamics, and can place stationary solitons close to breathing instabilities while preserving a compact four-coordinate description \cite{karpovRamanSelfFrequencyShift2016,wangStimulatedRamanScattering2018,milianSolitonsFrequencyCombs2015,semaanTemporalSolitonGeneration2026}. 
The Raman-extended LLE therefore provides a concrete deterministic setting in which realistic noise-routing channels are already present, while remaining simple enough for analytic stochastic reduction. 

Existing work has largely progressed along two distinct directions.
On one side, variational methods have long been used to reduce soliton dynamics to a small number of physically interpretable parameters, such as amplitude, width, frequency shift, position, chirp, or phase, depending on the system and perturbation considered. 
In weakly perturbed propagation settings, this line of work is represented by variational treatments of conservative or weakly perturbed solitons \cite{BondesonLisakAnderson1979,Anderson1983,KivsharMalomed1989,hasegawaOpticalSolitonsFibers2003}. 
In driven-dissipative cavity systems, related reductions have been developed for driven-damped nonlinear Schr\"odinger dynamics and for cavity solitons subject to phase modulation, focused or pulsed driving, trapping potentials, and parametric driving \cite{nozakiLowdimensionalChaosDriven1986a,sahooStabilityVariationalAnalysis2019,sahaVariationalApproachStudy2019,cardosoLocalizedSolutionsLugiatoLefever2017,liEfficiencyPulsePumped2022a,masarabiDepletionlimitedKerrSolitons2023,rossiTemporalTweezing2024,englebertDynamicsDrivenDissipative2026a}. 
Related collective-coordinate ideas have also been extended to stochastic partial differential equations \cite{Cartwright2019SPDE}.
On the other side, noise studies of Kerr solitons and microcombs have addressed timing jitter and quantum diffusion \cite{matskoTimingJitterMode2013,baoQuantumDiffusionMicrocavity2021,Seibold2022QuantumDynamicsPRA,jinSelfsuppressedQuantumlimitedTiming2024}, field- or mode-level fluctuation correlations, multimode squeezing, and stochastic phase-noise theory \cite{chemboFluctuationsCorrelationsKerr2020,liuStochasticApproachPhase2023,guidryMultimodeSqueezingSoliton2023}, as well as specific conversion or suppression mechanisms such as spectral purification, dispersive-wave-induced limits, linewidth reshaping, and all-optical quenching \cite{Weng2019SpectralPurification,yangDispersivewaveInducedNoise2021,leiOpticalLinewidthSoliton2022,Moille2025AllOptical,Shandilya2025Azimuthal,weiKerrinducedNoiseQuenching2026}. 
What is still missing, however, is a framework that combines these two perspectives: a stochastic reduced description in which field-level noise is projected onto the core soliton coordinates, and the resulting observable spectra are resolved into direct stochastic injection and deterministic transfer among amplitude, frequency, timing, and phase.

Here we develop a pathway-resolved stochastic collective-coordinate theory for stationary driven-dissipative solitons, using the generalized Lugiato--Lefever equation with Raman response as a representative model. As summarized in Fig.~\ref{fig:toy_overview}, field-level vacuum fluctuations are projected onto four collective coordinates: amplitude, frequency shift, temporal position, and global phase. Linearizing the resulting reduced Langevin model about a stable stationary soliton yields an analytic power-spectral-density matrix that separates direct stochastic injection from deterministic transfer among coordinates. This decomposition reveals how measurable intensity, timing, and phase-noise spectra are assembled from distinct internal pathways, including Gordon--Haus-type frequency-to-timing conversion, amplitude-to-phase transfer, and Raman-enabled cascaded channels. It also identifies the low-detuning spectral hump as a driven response of an underdamped amplitude--phase subsystem preceding the breathing instability. 
Agreement with stochastic simulations of both the reduced model and the full generalized Lugiato--Lefever equation confirms the validity of the reduction across most of the stable stationary single-soliton regime. Finally, we connect the collective-coordinate spectra to experimentally accessible observables.

The paper is organized as follows. Section~\ref{sec:model_to_normalization} introduces the model and reduced soliton description. Section~\ref{sec:stochastic_theory} develops the linearized stochastic theory. Section~\ref{sec:pathway_spectra} presents the pathway-resolved spectra, benchmarks them against stochastic simulations, and connects to experimentally relevant observables.
Section~\ref{sec_conclusions} presents our conclusions.

\section{Physical model, normalization, and soliton parametrization}
\label{sec:model_to_normalization}

This section develops the deterministic foundation of the reduced theory. We introduce the generalized mean-field model in dimensional and normalized form, define the four-coordinate soliton ansatz and its mapping to physical observables, and summarize representative normalization scales for relevant resonator platforms. We then derive the deterministic collective-coordinate equations via the Lagrange reduction, analyze their stationary solutions, and benchmark them against stationary solitons of the full LLE to delimit the regime where the reduced description is accurate.

\subsection{Generalized mean-field model and normalization}
\label{ssec:physical_model}

{We model a coherently driven Kerr resonator with a generalized LLE that includes optional walk-off, linearized Raman response, and a field-level Langevin source. Dimensional quantities are written with a tilde, whereas normalized quantities are written without a tilde.}

The dimensional evolution equation is
\begin{equation}
\label{eq:LLE_phys}
\begin{aligned}
\frac{\partial \tilde A}{\partial \tilde t}
  &= \underbrace{-\frac{\alpha}{2}\, \tilde A
     - i\, \tilde\delta_\omega\, \tilde A}_{\text{loss and detuning}}
     \underbrace{- \frac{\Delta t}{t_\mathrm{r}}\, \frac{\partial \tilde A}{\partial \tilde\tau}}_{\text{moving-frame correction}}
     \underbrace{- i\, \frac{\beta_2}{2\beta_1}\, 
         \frac{\partial^2 \tilde A}{\partial \tilde\tau^{2}}
        }_{\text{dispersion}}\\
  &\quad
     + \underbrace{i\, \frac{\gamma}{\beta_1}\!\Bigl(
        \lvert\tilde A\rvert^{2}-\tau_\mathrm{R}\, 
        \partial_{\tilde\tau}\lvert\tilde A\rvert^{2}\Bigr)\tilde A}_{\text{Kerr + Raman}}
     + \underbrace{\frac{\sqrt{\theta\tilde P_{\rm in}}}{t_\mathrm{r}}}_{\text{coherent drive}}
     + \underbrace{\tilde F(\tilde t, \tilde\tau)}_{\text{Langevin noise}} .
\end{aligned}
\end{equation}

Here $\tilde A(\tilde t,\tilde\tau)$ denotes the intracavity field envelope, where $\tilde t$ is the slow time describing roundtrip-to-roundtrip evolution and $\tilde\tau$ is the retarded fast-time coordinate. The round-trip time is $t_\mathrm{r}=\beta_1 L$, with $L$ the cavity round-trip length.
The free spectral range (FSR) is $f_\mathrm{r}=1/t_\mathrm{r}$, and the corresponding angular mode spacing is $\Omega_\mathrm{r}=2\pi f_\mathrm{r}$. We take $\alpha$ to be the energy-decay rate, and the corresponding roundtrip energy loss is approximately
\begin{equation}
\alpha_\mathrm{r}\simeq \alpha t_\mathrm{r}.
\end{equation}
The detuning $\tilde\delta_\omega$ is the angular-frequency offset between the driving laser frequency $\omega_0$ and one relevant cavity resonance.
The notation entering Eq.~\eqref{eq:LLE_phys} is summarized in Table~\ref{table_notation}.

\begin{table}[h!]
\centering
\setlength{\tabcolsep}{0.5pt}
\renewcommand{\arraystretch}{1.2}
\begin{tabularx}{86mm}{|c|Y|c|Y|}
\hline
\textbf{} & \multicolumn{1}{|c|}{\textbf{Meaning}} & \textbf{} & \multicolumn{1}{|c|}{\textbf{Meaning}} \\
\hline
$\tilde t$ & slow (evolution) time &
$\tilde\tau$ & fast (retarded) time \\
\hline
$t_\mathrm{r}$ & roundtrip time &
$\Omega_\mathrm{r}$ & angular mode spacing, $\Omega_\mathrm{r}=2\pi f_\mathrm{r}=2\pi/t_\mathrm{r}$ \\
\hline
$\beta_1$ & inverse group velocity, $\beta_1=t_\mathrm{r}/ L$ &
$\alpha$ & energy decay rate; inverse of cavity photon lifetime \\
\hline
$\alpha_r$ & roundtrip energy loss (dimensionless) &
$\tilde\delta_\omega$ & detuning (angular frequency) \\
\hline
$\tilde\delta$ & detuning (roundtrip phase), $\tilde\delta_\omega=\tilde\delta/t_\mathrm{r}$ &
$\Delta t$ & frame delay adjust per roundtrip\\
\hline
$\beta_k$ & $k$-th order dispersion coefficient &
$\gamma$ & Kerr nonlinear coefficient of the guided mode \\
\hline
$\tau_R$ & effective Raman time constant (e.g. $\tau_R\approx3\,\rm fs$ for fiber) &
$\theta$ & coupling ratio (drive efficiency) \\
\hline
$\tilde P_{\rm in}$ & pump power &
$\tilde F$ & Langevin noise (field-level) \\
\hline
\end{tabularx}
\caption{Notation used in Eq.~\eqref{eq:LLE_phys}.}
\label{table_notation}
\end{table}

For the collective-coordinate analysis, it is convenient to normalize the equation so that the field decay sets the slow-time unit, the Kerr coefficient becomes unity, and second-order dispersion appears only through its sign. The characteristic scales for fast time, slow time, field amplitude, energy, frequency, and phase are, respectively,
\begin{equation}
\label{eq:scales}
\begin{aligned}
  \tau_\mathrm{s} &= \sqrt{\frac{|\beta_2|}{\alpha\, \beta_1}}, \quad
  &t_\mathrm{s} = \frac{2}{\alpha}, \quad
  &A_\mathrm{s}  = \sqrt{\frac{\alpha\, \beta_1}{2\gamma}}, \\
  E_\mathrm{s}  &= A_\mathrm{s}^{2}\tau_\mathrm{s}, \quad
  &f_\mathrm{s} = \frac{\alpha}{2}, \quad
  &\phi_\mathrm{s} = 1 .
\end{aligned}
\end{equation}

Since Eq.~\eqref{eq:LLE_phys} contains the linear field-loss term $-(\alpha/2)\tilde A$, the slow-time scale $t_\mathrm{s}=2/\alpha$ corresponds to the field-decay time, while the associated intracavity energy lifetime is $1/\alpha$. Introducing the normalized variables
\begin{equation}
\begin{aligned}
  t   &= \frac{\tilde t}{t_\mathrm{s}},
&
  \tau &= \frac{\tilde\tau}{\tau_\mathrm{s}},
&
  A   &= \frac{\tilde A}{A_\mathrm{s}},
&
  E   &= \frac{\tilde E}{E_\mathrm{s}},
\end{aligned}\label{eq_scaling_varibles}
\end{equation}
where $\tilde E$ and $E$ denote the dimensional and dimensionless soliton energies, respectively (see the next section), Eq.~\eqref{eq:LLE_phys} becomes
\begin{equation}
\label{eq:normalized}
\begin{aligned}
  \frac{\partial A}{\partial t}
  =& \Bigl[
        -1 - i\delta
        - d_1\, \partial_\tau
        + i\, \partial_{\tau}^{2}
     \Bigr]A \\
    &+ i\!\Bigl(\lvert A\rvert^{2}
               - d_\mathrm{R}\, \partial_\tau\lvert A\rvert^{2}\Bigr)A
    + P + F(t,\tau),
\end{aligned}
\end{equation}
with dimensionless parameters
\begin{equation}
\label{eq_normalized_parameters}
\begin{aligned}
  d_1  &= \frac{2\, \Delta t}{\alpha\, t_\mathrm{r}\, \tau_\mathrm{s}},
&
   \delta &= \frac{2\tilde\delta_\omega}{\alpha}
         = \frac{2\tilde\delta}{\alpha_\mathrm{r}}, 
&
 &
\\
  d_\mathrm{R}  &= \frac{\tau_\mathrm{R}}{\tau_\mathrm{s}},
&
  P    &= \frac{2\, \sqrt{\theta\tilde P_{\rm in}}}{\alpha\, t_\mathrm{r}\, A_\mathrm{s}},
&
  F(t,\tau) &= \frac{2\, \tilde F(\tilde t,\tilde\tau)}{\alpha\, A_\mathrm{s}}.
\end{aligned}
\end{equation}

\begin{table*}[!ht]
\centering
\caption{
Normalization scales for representative Kerr-resonator platforms. The table shows how the same dimensionless stationary solutions and normalized fluctuation spectra map onto different physical pulse and noise scales in microresonators and fiber-loop resonators. The $Q$ factors are evaluated at $\lambda_0=1550\,\mathrm{nm}$.
}
\label{tab:scales}
\small
\setlength{\tabcolsep}{5pt}
\begin{tabular}{lcccccccccc}
\toprule
& \multicolumn{5}{c}{Resonator parameters} & \multicolumn{5}{c}{Normalization scales} \\
\cmidrule(lr){2-5}\cmidrule(lr){6-11}
Platform & FSR & $\gamma$ [W$^{-1}$m$^{-1}$] & Finesse & $Q$-factor & $t_\mathrm{r}$ & $f_\mathrm{s}$ & $t_\mathrm{s}$ & $\tau_\mathrm{s}$ [ps] & $A_\mathrm{s}^2$ [W] & $E_\mathrm{s}$ [pJ] \\
\midrule
Silica wedge\cite{baoQuantumDiffusionMicrocavity2021} & $24.41\,\mathrm{GHz}$ & $2.7\times10^{-3}$ & 11626 & $9.21\times10^{7}$ & $40.96\,\mathrm{ps}$ & $6.60\,\mathrm{MHz}$ & $151.6\,\mathrm{ns}$ & 0.619 & 10.647 & 6.586 \\
Si$_3$N$_4$ ring\cite{leiOpticalLinewidthSoliton2022} & $227.5\,\mathrm{GHz}$ & $0.9$ & 9409 & $8.0\times10^{6}$ & $4.396\,\mathrm{ps}$ & $75.95\,\mathrm{MHz}$ & $13.17\,\mathrm{ns}$ & 0.274 & 0.591 & 0.162 \\
Silica disk\cite{jinSelfsuppressedQuantumlimitedTiming2024} & $11.4\,\mathrm{GHz}$ & $3.56\times10^{-4}$ & 44206 & $7.5\times10^{8}$ & $87.72\,\mathrm{ps}$ & $0.810\,\mathrm{MHz}$ & $1.234\,\mu\mathrm{s}$ & 1.260 & 21.181 & 26.687 \\
MgF$_2$ WGM\cite{herrTemporalSolitonsOptical2014} & $35.2\,\mathrm{GHz}$ & $4.1\times10^{-4}$ & 78200 & $4.0\times10^{8}$ & $28.41\,\mathrm{ps}$ & $1.414\,\mathrm{MHz}$ & $707.4\,\mathrm{ns}$ & 0.853 & 15.754 & 13.431 \\
\midrule
Fiber ring\cite{leoTemporalCavitySolitons2010} & $0.539\,\mathrm{MHz}$ & $1.8\times10^{-3}$ & 24 & $8.61\times10^{9}$ & $1854\,\mathrm{ns}$ & $70.1\,\mathrm{kHz}$ & $14.26\,\mu\mathrm{s}$ & 5.41 & 0.190 & 1.03 \\
Fiber ring\cite{englebertTemporalSolitonsCoherently2021} & $3.969\,\mathrm{MHz}$ & $1.0\times10^{-3}$ & 265 & $1.29\times10^{10}$ & $252\,\mathrm{ns}$ & $47.1\,\mathrm{kHz}$ & $21.2\,\mu\mathrm{s}$ & 6.52 & 0.236 & 1.54 \\
\bottomrule
\end{tabular}
\end{table*}

Here the second-order dispersion term is written as \(+\,i\,\partial_\tau^2\), corresponding to the anomalous-dispersion case \(\beta_2<0\), since bright temporal cavity solitons exist only in this regime. 
The cavity finesse and the corresponding quality factor read
\begin{equation}
\label{eq:finesse_relation}
\begin{aligned}
\mathcal{F}
&=\frac{2\pi}{\alpha_\mathrm{r}}
=\frac{\pi t_\mathrm{s}}{t_\mathrm{r}}
=\frac{\Omega_\mathrm{r}}{2f_\mathrm{s}},\\
Q &= \frac{\omega_0}{\Omega_\mathrm{r}}\,\mathcal{F}=\frac{\omega_0}{2f_\mathrm{s}}.
\end{aligned}
\end{equation}

For comparison with the notation commonly used in the microresonator-comb
literature, Eq.~\eqref{eq:LLE_phys} can also be written in terms of the integrated-dispersion parameters \(D_k\) and the single-photon Kerr shift \(g\). With the azimuthal coordinate \(\varphi=D_1\tilde\tau\), the angular-domain field \(\mathcal A(\tilde t,\varphi)=
\sqrt{t_\mathrm{r}/(\hbar\omega_0)}\,
\tilde A\) is normalized such that \((2\pi)^{-1}\int_0^{2\pi}|\mathcal A|^2\,d\varphi\) represents the
intracavity photon number. Equation~\eqref{eq:LLE_phys} becomes 
\begin{equation}
\label{eq:LLE_modal_full}
\begin{aligned}
\frac{\partial \mathcal{A}}{\partial \tilde t}
=&
-\left(\frac{\alpha}{2}+i\tilde\delta_\omega\right)\mathcal A
-\frac{\Delta t}{t_\mathrm{r}}D_1
\frac{\partial\mathcal A}{\partial\varphi}
+i\frac{D_2}{2}
\frac{\partial^2\mathcal A}{\partial\varphi^2}
\\
&+
i g
\left(
|\mathcal A|^2
-
D_1\tau_\mathrm{R}
\frac{\partial |\mathcal A|^2}{\partial\varphi}
\right)\mathcal A
+
\sqrt{\kappa_\mathrm{ext}}\,s_\mathrm{in}
+
F_{\mathcal A}.
\end{aligned}
\end{equation}
with
\(
D_1=\Omega_\mathrm{r},
D_2=-\frac{\beta_2 D_1^2}{\beta_1}\),
\(
g=\frac{\gamma\hbar\omega_0 D_1}{2\pi\beta_1}\),
\(
s_\mathrm{in}=\sqrt{\frac{\tilde P_\mathrm{in}}{\hbar\omega_0}}\), and 
\(\kappa_\mathrm{ext}=\frac{\theta}{t_\mathrm{r}}, 
\) and
\(
F_\mathcal{A}=\sqrt{\frac{t_{\rm r}}{\hbar\omega_0}}
\tilde F
\).
Note that Eq.~\eqref{eq:LLE_modal_full} is not a separate model, but a
dictionary between the waveguide-style dimensional variables used in
Eq.~\eqref{eq:LLE_phys} and the microcomb convention; after applying
the scalings above, it leads to the same normalized LLE form used in the
remainder of this work.

\subsection{Soliton ansatz and parameter mapping}
\label{ssec:ansatz_mapping}

We next specify the soliton ansatz that underlies the collective-coordinate reduction. The central approximation is that the intracavity field is represented by a localized soliton core whose evolution is captured by a small set of slowly varying collective coordinates. In the present reduced description, these are the amplitude $\eta$, frequency shift $\Omega$, temporal center $\tau_0$, and global phase $\phi$. In normalized variables, the soliton core is written as
\begin{equation}
\label{eq:nor_ansatz}
A(t,\tau)
=
\sqrt{2}\,\eta\,
\sech\!\bigl[\eta(\tau-\tau_0)\bigr]\,
e^{-i\Omega(\tau-\tau_0)+i\phi}.
\end{equation}
The corresponding dimensional form is
\begin{equation}
\label{eq:_ansatz}
\tilde A(\tilde t,\tilde\tau)
=
\tilde A_0\,
\sech\!\Bigl(\frac{\tilde\tau-\tilde\tau_0}{\tilde\tau_\mathrm{w}}\Bigr)\,
e^{-i\tilde\Omega(\tilde\tau-\tilde\tau_0)+i\tilde\phi}.
\end{equation}

Within this ansatz family, the single parameter $\eta$ determines the basic soliton observables in normalized units:
\begin{equation}
\label{eq_eta_to_pulse_paras}
A_0=\sqrt{2}\eta,
\qquad
\tau_\mathrm{w}=\eta^{-1},
\qquad
E=\int_{-\infty}^{\infty}|A|^2\,\dd\tau=4\eta.
\end{equation}
Thus, the reduced description is naturally soliton-centered: $\eta$ determines the core amplitude, width, and energy, while $\Omega$, $\tau_0$, and $\phi$ describe the spectral shift, temporal position, and common phase.


The normalized and dimensional parameters are related through the scaling factors introduced in Eq.~\eqref{eq:scales}:
\begin{alignat}{3}
\tilde A_0 &= \sqrt{2}\eta\,A_\mathrm{s},
&\qquad
\tilde\tau_\mathrm{w} &= \frac{\tau_\mathrm{s}}{\eta},
&\qquad
\tilde E &= 4\eta E_\mathrm{s}, \label{eq:nor_units_tau_w}\\
\tilde\tau_0 &= \tau_0\,\tau_\mathrm{s},
&\qquad
\tilde\Omega &= \frac{\Omega}{\tau_\mathrm{s}},
&\qquad
\tilde\phi(\tilde t) &= \phi(t).
\end{alignat}
Eliminating \(\eta\) in Eqs.\eqref{eq:nor_units_tau_w} gives the relations
\begin{equation}
\label{eq:duration_energy_relations}
\tilde\tau_\mathrm{w}
=
\frac{4E_\mathrm{s}\tau_\mathrm{s}}{\tilde E},
\qquad
\tilde E
=
\frac{4E_\mathrm{s}\tau_\mathrm{s}}{\tilde\tau_\mathrm{w}},
\qquad
|\tilde A_0|^2
=
2A_\mathrm{s}^2
\Bigl(\frac{\tau_\mathrm{s}}{\tilde\tau_\mathrm{w}}\Bigr)^2.
\end{equation}
These identities, together with Eq.~\eqref{eq:scales}, will be used below when the noise spectra are recast in terms of experimentally accessible pulse parameters.

It is also useful to anticipate the large-detuning stationary regime discussed below in Sec.~\ref{ssec:stationary_soliton}. 
There, one has the asymptotic relation \(\eta \simeq \sqrt{\delta}\), which, together with the scale dictionary introduced above, yields the approximate physical estimates
\begin{equation}
|\tilde A_0|^2 \simeq 2A_\mathrm{s}^2\,\delta,
\qquad
\tilde\tau_\mathrm{w} \simeq \frac{\tau_\mathrm{s}}{\sqrt{\delta}},
\qquad
\tilde E \simeq 4E_\mathrm{s}\sqrt{\delta}.
\end{equation}
These expressions are useful as back-of-the-envelope estimates: once the scaling parameters of a resonator are known, they provide immediate estimates of the peak power, pulse duration, and pulse energy associated with the large-detuning solitons that the platform can support.

\subsection{Representative physical scales}
\label{ssec:representative_scales}

The normalized LLE and the corresponding reduced collective-coordinate model are universal only at the dimensionless level. Any comparison with experiment therefore requires restoring the platform-dependent scales that convert normalized soliton parameters and fluctuation spectra into laboratory units. Table~\ref{tab:scales} summarizes these normalization scales for representative Kerr-resonator platforms.

Because both microresonators and fiber-loop resonators reduce to the same normalized LLE form \cite{Coen2013Scaling}, the same dimensionless stationary solutions can correspond to very different physical pulse durations, peak powers, slow-time relaxation rates, and noise levels. In particular, $t_\mathrm{s}$ sets the slow-time scale of the reduced dynamics and the frequency-axis conversion of the spectra, $\tau_\mathrm{s}$ converts normalized pulse widths, timing shifts, and frequency shifts into physical units, and $(A_\mathrm{s},E_\mathrm{s})$ set the characteristic scales of peak power and pulse energy. Once these quantities are specified for a given resonator, the normalized stationary states and PSDs can be translated directly into experimentally relevant observables.

Among the entries in Table~\ref{tab:scales}, we will later use the fiber-loop resonator of Ref.~\cite{leoTemporalCavitySolitons2010} and the silica-wedge microresonator of Ref.~\cite{baoQuantumDiffusionMicrocavity2021} as two illustrative examples when converting the normalized theory to physical units.
In gain-assisted resonators, these scales can be reconfigured by Raman-gain-induced changes of the effective loss \cite{semaanTemporalSolitonGeneration2026}.

\subsection{Deterministic reduced equations}
\label{ssec:deterministic_reduced}

Here we introduce a deterministic reduced model for the soliton collective coordinates using an averaged Lagrangian for the conservative core together with a Lagrange--d'Alembert treatment of the nonconservative terms. This reduction strategy builds on variational and collective-coordinate approaches developed for conservative soliton systems \cite{BondesonLisakAnderson1979,Anderson1983,KivsharMalomed1989,hasegawaOpticalSolitonsFibers2003} and later extended to driven cavity-soliton settings governed by the LLE \cite{sahooStabilityVariationalAnalysis2019,sahaVariationalApproachStudy2019,cardosoLocalizedSolutionsLugiatoLefever2017,liEfficiencyPulsePumped2022a,masarabiDepletionlimitedKerrSolitons2023,rossiTemporalTweezing2024,englebertDynamicsDrivenDissipative2026a}. 
A distinctive feature of the present derived result is a refined stationary phase-locking. 
Compared with earlier reduced descriptions, this correction improves the agreement of the reduced stationary branch with generalized LLEs and, crucially, provides a more accurate fixed point for the subsequent stochastic reduction and noise analysis.

Applying this reduction to the ansatz of Sec.~\ref{ssec:ansatz_mapping} yields coupled evolution equations for the four collective coordinates
\(
\eta,\Omega,\tau_0,\phi
\) [see Fig.~\ref{fig:toy_overview}],
{which we collect as}
\begin{equation}
\label{eq_SODEs}
  \mathbf{x}(t)
  =
  \begin{pmatrix}
    \eta(t),\,
    \Omega(t),\,
    \tau_0(t),\,
    \phi(t)
  \end{pmatrix}^{\!\top}.
\end{equation}

The reduced dynamics can be written schematically as
\begin{equation}
\label{eq:param_odes}
  \frac{\dd \mathbf{x}}{\dd t}
  = \mathbf{G}(\mathbf{x}) + \mathbf{F}(t),
\end{equation}
where $\mathbf{F}(t)=(F_\eta, F_\Omega, F_{\tau_0}, F_\phi)^{\!\top}$ are the \emph{projected} (real-valued) noise forces and
$\mathbf{G}=(G_\eta, G_\Omega, G_{\tau_0}, G_\phi)^{\!\top}$ collects the deterministic drift terms.
In this section we first focus on the deterministic backbone $\dd\mathbf{x}/\dd t=\mathbf{G}(\mathbf{x})$ and later return to
$\mathbf{F}(t)$ when analyzing noise.

The drift terms are [see derivation in Sec.~1 in the Supplemental Document]
\begin{subequations}
\label{eq:G_components}
\begin{align}
  G_\eta
  &=
  \frac{\pi P}{\sqrt{2}}\mathcal{S}\cos\phi
  - 2\eta,
  \\
  G_\Omega
  &=
  -\frac{16}{15}\, d_{R}\, \eta^{4}
  - \sqrt{2}\, P\,R\, \mathcal{S}\cos\phi,
  \\
  G_{\tau_0}
  &=
  d_1 - 2\Omega
  + \frac{\sqrt{2}\, \pi^{2}P}{4\eta^{2}}\,\mathcal{S}\mathcal{T}\sin\phi,
  \label{eq_G_tau0}
  \\
  G_\phi
  &=
  \Omega^{2} + \eta^{2} - \delta
  - \frac{3\pi P}{4\sqrt2\, \eta}\,\mathcal{S}\sin\phi(1+\frac{2}{3}R\mathcal{T}).
\end{align}
\end{subequations}

Here $G_\eta$ expresses the balance between drive and linear loss, $G_\phi$ provides the phase-locking condition between the pump and the soliton core, $G_{\tau_0}$ describes the kinematic relation between walk-off and frequency-shift, and $G_\Omega$ contains both the Raman-induced redshift and a smaller drive-induced correction.

The overlap factors
\(
\mathcal{S}=\sech(R)
\)
{and}
\(
\mathcal{T}=\tanh(R)
\)
arise from the projection integrals. The parameter
\begin{equation}
\label{eq_R_omega}
  R = \frac{\Omega}{w_\Omega} = \frac{\pi \Omega}{2\eta}
\end{equation}
measures the redshift in units of the soliton spectral width $w_\Omega=2\eta/\pi$. 
For simplicity, without losing generality, we assume 
\begin{equation}
|R|\ll1,
\label{eq_small_shift_R_ll_1}
\end{equation}
so the redshift is small compared with the soliton bandwidth.
In terms of real units, this amounts to the condition
\begin{equation}
    \tilde{\tau}_{\rm w} \gg \sqrt[3]{\tau_{\rm R}\tau_{\rm s}^2}.
\end{equation}
Accordingly, as $\tilde{\tau}_{\rm w}$ decreases toward the characteristic scale $\sqrt[3]{\tau_{\rm R}\tau_{\rm s}^2}$, the parameter $R$ grows and the small-$R$ approximation becomes progressively less accurate.

Expanding $\sech(R)=1+\mathcal{O}(R^2)$ and $\tanh(R)=\mathcal{O}(R)$, and retaining only leading-order terms gives
\begin{subequations}
\label{eq:G_components_simplified}
\begin{align}
G_\eta &= -2\eta+\frac{\pi P}{\sqrt2}\cos\phi, \\
G_\Omega &= -\frac{\pi P}{\sqrt2\eta}\Omega\cos\phi-\frac{16}{15}d_\mathrm{R}\eta^4, \\
G_{\tau_0} &= d_1-2\Omega \\
G_\phi &= -\delta+\eta^2+\Omega^2-\frac{3\pi P}{4\sqrt2\eta}\sin\phi.
\end{align}
\end{subequations}
These leading-order equations retain the couplings that control both the stationary fixed point and the linearized fluctuation-transfer problem discussed later. 

\subsection{Stationary soliton relations}
\label{ssec:stationary_soliton}

\begin{figure*}[!t]
  \centering
  \includegraphics[width=1\linewidth]{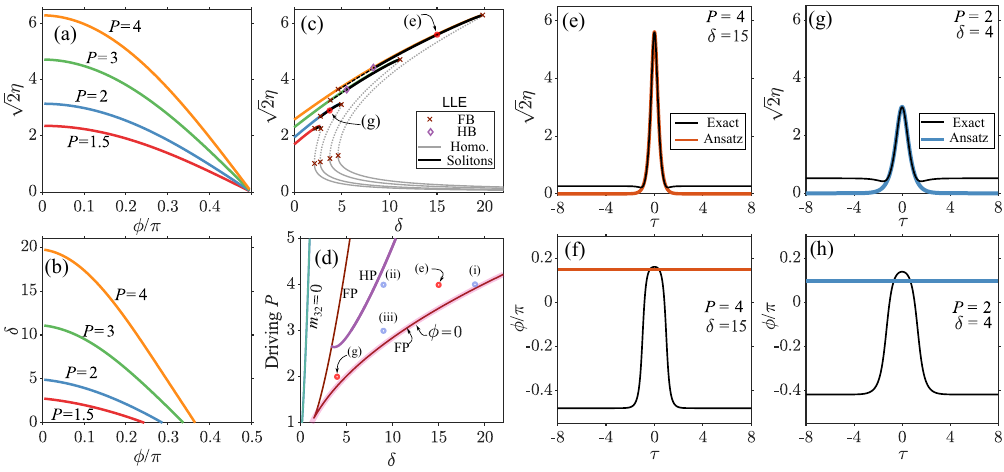}
  \caption{Stationary reduced-model parameters (Eq.~\eqref{eq:stationary}) and comparison with stationary solutions of the full mean-field equation (Eq.~\eqref{eq:normalized}). (a,b) Dependence of the reduced-model amplitude $\sqrt{2}\eta$ and detuning $\delta$ on the stationary phase $\phi$ for several pump values $P$, when $\Omega=0$. (c) $\sqrt{2}\eta$ versus $\delta$: reduced-model curves (colored) compared with stationary solitons of Eq.~\eqref{eq:normalized} (black; solid stable, dashed unstable); the homogeneous steady states are plotted by solid gray curves. Fold (FB) and Hopf (HB) bifurcations from the LLE are indicated. (d) Organization of the soliton region in the $(\delta,P)$ plane, showing the fold/Hopf curves of the full LLE together with the reduced-model loci $\phi=0$. (e--h) Representative temporal amplitude and phase profiles comparing stationary LLE solitons (black) with the ansatz core (colored).
  }
  \label{fig:detuning_vs_phase}
\end{figure*}

Stationary solitons correspond to fixed points $\mathbf{x}_{\mathrm{s}}$ of the reduced dynamical system, defined by $\mathbf{G}(\mathbf{x}_{\mathrm{s}})=\mathbf{0}$. 
Eq.~\eqref{eq:G_components_simplified} gives
\begin{subequations}
\label{eq:stationary}
\begin{align}
\eta
  &= \frac{\pi P}{2\sqrt{2}}\cos\phi,
  \label{eq:eta_stat}\\
\Omega
  &= -\frac{8}{15}\, d_{R}\, \eta^{4},
  \label{eq:Omega_stat}\\
d_1
  &= 2\Omega,
  \label{eq:d1_stat}\\
\delta
  &= \eta^{2}+\Omega^{2}-\frac{3}{2}\tan\phi.
  \label{eq:delta_stat}
\end{align}
\end{subequations}

These relations summarize the structure of the stationary state. 
Equation~\eqref{eq:Omega_stat} gives the leading-order Raman-induced soliton redshift, while Eq.~\eqref{eq:d1_stat} fixes the corresponding group-delay, or comoving-frame, condition associated with this redshift. 
Equation~\eqref{eq:eta_stat} shows that the soliton phase $\phi$ mediates the balance between the external drive and the soliton core amplitude, whereas Eq.~\eqref{eq:delta_stat} gives the associated phase-locking condition involving detuning, phase, frequency shift, and amplitude. 
Eliminating $\phi$ between Eqs.~\eqref{eq:eta_stat} and \eqref{eq:delta_stat} yields the amplitude--detuning relation
\begin{equation}
\label{eq:master_exact}
\pi^{2}P^{2}
=
8\,\eta^{2}
\left[
1+\frac{4}{9}\bigl(\delta-\eta^{2}-\Omega^{2}\bigr)^{2}
\right].
\end{equation}
This equation is cubic in \(u=\eta^{2}\). In the regime where it admits three real roots, the physically relevant branch is the largest positive one, which yields
\begin{equation}
\eta = \sqrt{ \frac{2}{3} \left[ (\delta - \Omega^{2}) + \sqrt{ (\delta - \Omega^{2})^{2} - \frac{27}{4} } \cos( \frac{\arccos(\chi)}{3} ) \right] }
\end{equation}
with
\begin{equation}
\chi(P,\delta) = \frac{ 243\pi^{2}P^{2} - 64(\delta - \Omega^{2})^{3} - 1296(\delta - \Omega^{2}) }{ \left( 16(\delta - \Omega^{2})^{2} - 108 \right)^{3/2} }
\end{equation}
valid for \((\delta - \Omega^{2})^{2} > 27/4\) and \(|\chi| \le 1\).
Eq.~\eqref{eq:master_exact} only has one real solution when \((\delta - \Omega^{2})^{2} \leq 27/4\).

To compare the stationary reduced description against the full LLE, Fig.~\ref{fig:detuning_vs_phase} considers the weak redshift regime, $\Omega \simeq 0$, where the stationary states can be parameterized conveniently by the phase $\phi$. For each driving value $P$, increasing $\phi$ from $0$ decreases both the reduced-model amplitude $\sqrt{2}\eta$ [Fig.~\ref{fig:detuning_vs_phase}(a)] and the corresponding detuning $\delta$ [Fig.~\ref{fig:detuning_vs_phase}(b)]. Thus, $\phi=0$ corresponds to the largest detuning at the end of the branch, while larger $\phi$ moves the solution toward lower detuning. Eliminating $\phi$ yields the reduced-model relation $\sqrt{2}\eta(\delta)$ shown in Fig.~\ref{fig:detuning_vs_phase}(c). In that panel, the solid and dashed black curves denote the exact LLE soliton peak-amplitude branches, the gray curves denote the homogeneous solution, and the fold and Hopf bifurcations are marked explicitly. The reduced model curves track the exact stationary soliton branch well over most of its extent, especially at moderate and large detuning, where the pulse is strongly localized and the single-soliton ansatz is most accurate. 
Figure~\ref{fig:detuning_vs_phase}(d) shows the same organization in the $(\delta,P)$ plane by comparing the fold (FP) and Hopf (HP) boundaries of the full LLE \cite{godeyStabilityAnalysisSpatiotemporal2014,sunDynamicsDissipativeStructures2023} with the corresponding reduced-model loci. In particular, the reduced boundary $\phi=0$ follows the right-hand FP line. The representative cases marked by (e,g) in Fig.~\ref{fig:detuning_vs_phase}(c) are shown in Figs.~\ref{fig:detuning_vs_phase}(e)--(h): at large detuning $(P,\delta)=(4,15)$, the ansatz yields a near-exact match of the amplitude and core phase; at lower detuning $(P,\delta)=(2,4)$, it remains structurally robust despite a marginal phase deviation. Ultimately, the consistency of these stationary results ensures a reliable foundation for the stochastic noise analysis that follows.

Furthermore, several useful approximate expressions can be derived from the leading-order stationary soliton theory.  
At fixed pump strength $P$, the largest detuning supported by the theory is obtained at $\phi=0$. 
By using Eq.~\eqref{eq_normalized_parameters}, Eq.~\eqref{eq:eta_stat} converted to dimensional units yields
\begin{equation}
\label{eq:delta_max_phys}
\tilde\delta_{\max}
= \frac{\pi^2\, \theta\, \gamma\, \tilde P_{\rm in}\, L}{2\alpha_\mathrm{r}^2}
= \frac{\theta\, \gamma\, \tilde P_{\rm in}\, L}{8}\, \mathcal{F}^2.
\end{equation} 
Combining this relation with Eq.~\eqref{eq:nor_units_tau_w} gives the lower-bound estimate for the soliton width at fixed pump power,
\begin{equation}
\label{eq:tau_fwhm_min}
\tilde\tau_{\mathrm{w,min}}
\simeq
\frac{1}{\mathcal{F}}\,
\sqrt{\frac{4|\beta_2|}{\gamma\, \theta \tilde P_{\rm in}}}.
\end{equation}
Similarly, Eq.~\eqref{eq:Omega_stat} gives the Raman redshift of a soliton,
\begin{equation}
\label{eq:redshift_estimate_phys}
\tilde\Omega
=
-\frac{4}{15\pi}
\frac{\tilde\tau_\mathrm{R}\gamma^2L}{|\beta_2|}
|\tilde A_{0}|^{4}\,\mathcal{F}.
\end{equation}
At fixed pump strength, \(|\tilde\Omega|\propto\eta^4\) is maximized at the
maximum-amplitude point of the reduced stationary branch, \(\phi=0\). Using
the corresponding leading-order relation \(\eta^2\simeq\delta\) gives a
detuning-based upper-bound scaling estimate
\begin{equation}
\label{eq:redshift_detuning_estimate_phys}
\tilde\Omega_{\rm max}
\simeq
-\frac{16}{15\pi}
\frac{\tilde\tau_\mathrm{R}}{|\beta_2|L}\,
\tilde\delta_{\max}^{2}\mathcal{F}.
\end{equation}
The omitted \(\Omega^2\) term in Eq.~\eqref{eq:delta_stat} gives a
higher-order correction to this weak-redshift estimate.

\section{Noise projection and linearized stochastic theory}
\label{sec:stochastic_theory}

This section develops the fixed-point stochastic theory for small fluctuations about a stable stationary single-soliton state of the reduced model. We first formulate the linearized stochastic dynamics in collective-coordinate space, which separates direct noise injection from deterministic inter-coordinate conversion. We then analyze the structure of the linear drift matrix, derive the reduced diffusion matrix by projecting field-level vacuum noise onto the collective coordinates, and finally obtain the corresponding PSD matrix and pathway decomposition.

We restrict the field-level fluctuations to quantum vacuum noise in the standard Markovian input--output limit~\cite{GardinerCollett1985,GardinerZollerQuantumNoise}. 
In this limit, the reservoir associated with cavity loss is broadband on the scale of both the cavity linewidth and the reduced slow-time dynamics, so that the vacuum Langevin force is delta-correlated in time. 
This choice is consistent with the local damping term in the mean-field equation and provides the minimal quantum-limited noise floor of the resonator. 
It also gives the cleanest setting for the present analysis: because the injected noise is spectrally flat, the frequency dependence of the resulting PSDs can be attributed to deterministic transfer within the reduced soliton dynamics, rather than to spectral structure already present in the noise source. 
More general sources, including amplified-spontaneous-emission noise, pump technical noise, and detuning fluctuations, can be included by adding or replacing the constant diffusion matrix with the appropriate, possibly frequency-dependent, noise covariance, but are left for future work.


\subsection{Fixed-point stochastic linearization}
\label{ssec:stochastic_linearization}

We begin by formulating the reduced stochastic dynamics near a linearly stable stationary single-soliton fixed point. At this stage, the diffusion matrix is introduced only abstractly; its explicit evaluation from the field-level Langevin forcing is deferred to the following subsections.

Let
\[
\mathbf{x}(t)=\mathbf{x}_{\mathrm{s}}+\Delta\mathbf{x}(t),
\qquad
\Delta\mathbf{x}
=
(\Delta\eta,\Delta\Omega,\Delta\tau_0,\Delta\phi)^{\!\top},
\]
where $\mathbf{x}_{\mathrm{s}}$ is a stationary solution satisfying
$\mathbf{G}(\mathbf{x}_{\mathrm{s}})=\mathbf{0}$.
After restoring the stochastic forcing and linearizing Eq.~\eqref{eq:param_odes} about $\mathbf{x}_{\mathrm{s}}$, the reduced fluctuations obey
\begin{equation}
\label{eq:linear_system}
\frac{d}{dt}\,\Delta\mathbf{x}
=
\mathbf{M}\,\Delta\mathbf{x}
+\mathbf{F}(t),
\end{equation}
with Jacobian
\[
\mathbf{M}
=
\left.\frac{\partial \mathbf{G}}{\partial \mathbf{x}}\right|_{\mathbf{x}_{\mathrm{s}}}.
\]

The projected stochastic forcing $\mathbf{F}(t)$ has zero mean and defines the reduced diffusion matrix $\mathbf{D}$ through
\begin{equation}
\label{eq:D_def}
\big\langle \mathbf{F}(t)\,\mathbf{F}^{\top}(t')\big\rangle
=
\mathbf{D}\,\delta_\mathrm{D}(t-t').
\end{equation}
where $\delta_\mathrm{D}(\cdot)$ denotes the Dirac delta distribution, and the matrix $\mathbf{D}$ specifies how noise is injected directly into the collective coordinates, while $\mathbf{M}$ governs how those fluctuations are subsequently redistributed by deterministic linear conversion.
The present formulation applies to \emph{fixed-point, slow-time} fluctuations of the reduced model. It does not describe breathing states, switching events, or regimes in which the reference solution itself is time dependent. 

We next examine the structure of the drift matrix $\mathbf{M}$, since it determines the internal routing of any injected noise.
To expose the dominant inter-coordinate conversion channels, we first linearize the original drift terms (Eqs.~(\ref{eq:param_odes}, \ref{eq:G_components})), then do the same small-$R$ approximation (Eq.~\eqref{eq_small_shift_R_ll_1}). In this limit, the Jacobian takes the sparse form [see derivation in Sec.~2 in the Supplemental Document]
\begin{equation}
\label{eq:jacobian_simple}
\mathbf{M}=
\begin{pmatrix}
-2 & 0 & 0 & m_{14} \\
m_{21} & -2 & 0 & 0 \\
0 & m_{32} & 0 & 0 \\
m_{41} & m_{42} & 0 & -3/2
\end{pmatrix}.
\end{equation}
Thus, the coordinates $(\eta,\Omega,\phi)$ are linearly damped, whereas the timing coordinate $\tau_0$ remains neutral at this order, reflecting translational invariance of the soliton center.

\begin{figure*}
    \centering
    \includegraphics[width=\textwidth]{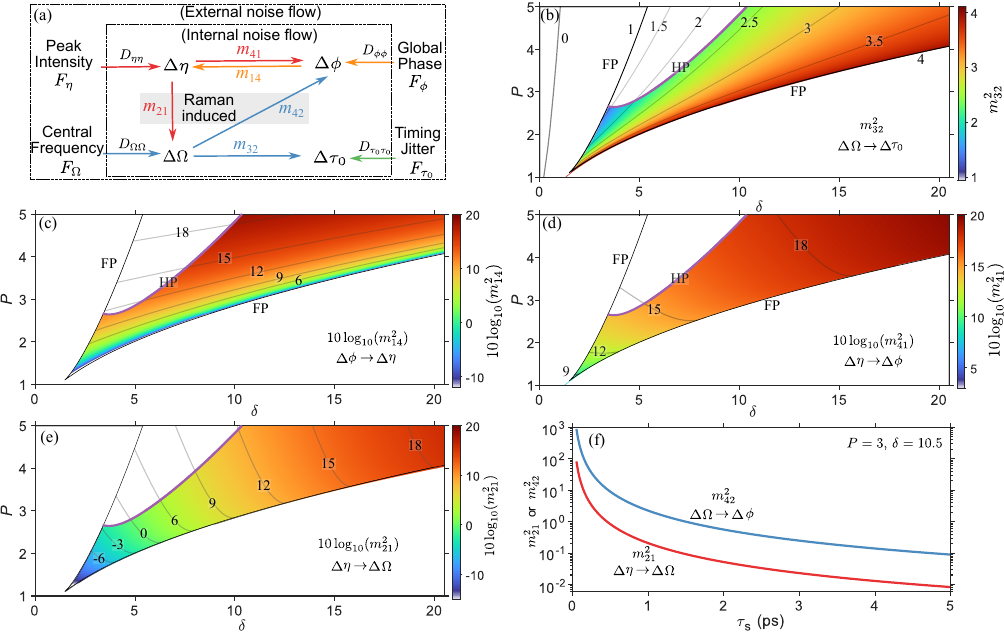}
    \caption{{Noise-transfer pathways and dominant linear couplings in the reduced soliton model. (a) Schematic of the fixed-point linearized stochastic dynamics, showing direct noise injection into the collective coordinates through the diagonal elements of $\mathbf{D}$ and deterministic conversion channels among $\Delta\eta$, $\Delta\Omega$, $\Delta\tau_{0}$, and $\Delta\phi$ generated by the Jacobian $\mathbf{M}$. (b--e) Maps of the dominant linear couplings $m_{32}^{2}$, $10\log_{10}(m_{14}^{2})$, $10\log_{10}(m_{41}^{2})$, and $10\log_{10}(m_{21}^{2})$ over the single-soliton existence region in the $(\delta,P)$ plane. (f) Raman-induced couplings $m_{21}^{2}$ and $m_{42}^{2}$ versus the fast-time scale $\tau_\mathrm{s}$ at $(P,\delta)=(3,\,10.5)$. FP denotes the fold boundary and HP the Hopf boundary.}}
    \label{fig:m_elements}
\end{figure*}

The nonzero off-diagonal coefficients are
\begin{align}
m_{14} &= -2\eta \tan\phi, \\
m_{21} &= -\frac{64}{15} d_\mathrm{R}\eta^3, \\
m_{32} &= \frac{\pi^2}{2\eta^2}\tan\phi - 2, \label{eq:m32_def}\\
m_{41} &= 2\eta + \frac{3\tan\phi}{2\eta}, \\
m_{42} &= -\frac{16}{15} d_\mathrm{R}\eta^4,
\end{align}
where $\tan\phi$ may be eliminated in favor of stationary quantities through Eq.~\eqref{eq:delta_stat} when convenient.

Figure~\ref{fig:m_elements}(a) summarizes the structure of Eq.~\eqref{eq:linear_system}. The outer channels represent direct stochastic injection into the reduced coordinates, quantified by the diagonal entries of $\mathbf{D}$, whereas the inner arrows represent deterministic conversion generated by the off-diagonal elements of $\mathbf{M}$. In this sense, panel (a) separates the noise-flow problem into \emph{external injection} and \emph{internal routing}. The latter is controlled entirely by the drift matrix and is therefore independent of the microscopic origin of the forcing.

\emph{Timing sector.}
The timing coordinate is distinguished by the fact that the third column of $\mathbf{M}$ vanishes identically: $\Delta\tau_0$ does not feed back into the other variables at linear order. This is the reduced-model expression of translational neutrality of the soliton core. Accordingly, within the present linearized description, timing fluctuations arise either from direct forcing in the $\tau_0$ coordinate or from conversion of frequency fluctuations through $m_{32}$. Figure~\ref{fig:m_elements}(b) shows that throughout most of the single-soliton region one typically has $1\lesssim m_{32}^2\lesssim 4$, so this conversion channel remains substantial. 

\emph{Amplitude--phase interconversion.}
The pair $(m_{14},m_{41})$ governs bidirectional coupling between amplitude and global phase, but the two directions are strongly asymmetric. Phase-to-amplitude conversion is controlled by $m_{14}$ and is therefore suppressed when the stationary phase approaches zero, as occurs at larger detuning [Fig.~\ref{fig:m_elements}(c)]. By contrast, amplitude-to-phase conversion is set by $m_{41}$ and typically remains finite even when $\phi\simeq0$; indeed, it often grows with detuning through the increase of $\eta$ [Fig.~\ref{fig:m_elements}(d)]. This asymmetry already indicates at the deterministic linear level that phase fluctuations can be seeded efficiently by amplitude fluctuations even in regimes where the reverse pathway is weak.

\emph{Raman-induced channels.}
The Raman response opens additional conversion routes by coupling amplitude fluctuations into frequency through $m_{21}$ and frequency fluctuations into phase through $m_{42}$. It therefore breaks the more weakly coupled non-Raman structure and enables an amplitude$\to$frequency$\to$timing/phase cascade. The coefficient $m_{21}$ varies strongly across the single-soliton region because it scales with $\eta^3$ [Fig.~\ref{fig:m_elements}(e)]. Both $m_{21}$ and $m_{42}$ also inherit explicit resonator dependence through $d_\mathrm{R}=\tau_\mathrm{R}/\tau_\mathrm{s}$ [Eq.~\eqref{eq_normalized_parameters}]. Figure~\ref{fig:m_elements}(f) illustrates this dependence at fixed $(P,\delta)=(3,\,10.5)$: decreasing $\tau_\mathrm{s}$ strengthens both Raman-mediated couplings, with $m_{42}$ remaining the larger of the two over the range shown.


\subsection{Field-level vacuum noise and projected diffusion matrix}
\label{ssec:noise_projection}

We now derive the reduced diffusion matrix by projecting the field-level Langevin forcing onto the collective coordinates. As a baseline example, we take the input fluctuations to be quantum vacuum noise. The purpose of this subsection is to connect the microscopic field-noise normalization to the effective stochastic forcing $\mathbf{F}(t)$ appearing in Eq.~\eqref{eq:linear_system}.

At the field level, vacuum noise must be distinguished from a phenomenological commuting complex white-noise source because its second-order correlations are \emph{ordered}. In particular, we take the only nonvanishing vacuum correlation to be
\begin{equation}
\label{eq:noise_phys_vacuum}
  \bigl\langle \tilde F(\tilde t,\tilde\tau)\,
  \tilde F^{\dagger}(\tilde t',\tilde\tau')\bigr\rangle
  =
  N^2\,
  \delta_\mathrm{D}(\tilde t-\tilde t')\,
  \delta_\mathrm{D}(\tilde\tau-\tilde\tau'),
\end{equation}
where $N^2$ is the noise strength with unit $\rm J/s$. For vacuum input,
\begin{equation}
\label{eq:vacuum_ordering_phys}
\bigl\langle \tilde F^{\dagger}(\tilde t,\tilde\tau)\,
\tilde F(\tilde t',\tilde\tau')\bigr\rangle = 0,
\qquad
\bigl\langle \tilde F \tilde F \bigr\rangle
=
\bigl\langle \tilde F^\dagger \tilde F^\dagger \bigr\rangle
=0.
\end{equation}
Equation~\eqref{eq:vacuum_ordering_phys} should therefore be interpreted as an \emph{ordered} vacuum correlator rather than as the symmetric covariance of a classical commuting field.

A convenient convention is \cite{matskoTimingJitterMode2013,jinSelfsuppressedQuantumlimitedTiming2024}
\begin{equation}
\label{eq:F_def_vacuum}
\tilde F(\tilde t,\tilde\tau)
=
\sqrt{\hbar\omega_0\,\alpha}\;
\xi(\tilde t,\tilde\tau),
\end{equation}
where $\omega_0$ is the carrier frequency and $\xi$ is a circular complex white-noise field with unit delta correlation,
\begin{equation}
\label{eq:xi_norm}
\bigl\langle \xi(\tilde t,\tilde\tau)\,
\xi^{\dagger}(\tilde t',\tilde\tau')\bigr\rangle
=
\delta_\mathrm{D}(\tilde t-\tilde t')\,
\delta_\mathrm{D}(\tilde\tau-\tilde\tau').
\end{equation}
This gives
\begin{equation}
N^2=\hbar\omega_0\,\alpha.
\end{equation}

Using
\(
F(t,\tau)=\tfrac{2}{\alpha A_\mathrm{s}}\tilde F(\tilde t,\tilde\tau)
\)
in Eq.~\eqref{eq_normalized_parameters},
together with two time scale in Eqs.~\eqref{eq_scaling_varibles}, the normalized vacuum forcing satisfies
\begin{equation}
\label{eq:noise_norm_vacuum}
  \bigl\langle F(t,\tau)\,
  F^{\dagger}(t',\tau')\bigr\rangle
  =
  B\,
  \delta_\mathrm{D}(t-t')\,\delta_\mathrm{D}(\tau-\tau'),
\end{equation}
with
\begin{equation}
\label{eq:B_def}
B=\frac{2\hbar\omega_0}{E_\mathrm{s}},
\qquad
E_\mathrm{s}=A_\mathrm{s}^2\tau_\mathrm{s}.
\end{equation}
This coefficient fixes the strength of the normalized vacuum forcing before reduction to collective-coordinate space.

Field-level vacuum noise is then projected onto the tangent space of the soliton manifold, yielding stochastic equations for the collective coordinates [Sec.~3 in the Supplemental Document].
The projected Langevin forces introduced abstractly in Eq.~\eqref{eq:linear_system} are defined by
\begin{equation}
\label{eq:proj_force_def}
F_i(t)=\int_{-\infty}^{\infty}\Re\!\bigl\{\Psi_i^*(\tau)\,F(t,\tau)\bigr\}\dd\tau,
\quad i\in\{\eta,\Omega,\tau_0,\phi\},
\end{equation}
where the adjoint kernels are 
\begin{equation}
\label{eq:kernels}
\begin{aligned}
\Psi_\eta(\tau) &=-\frac{1}{2}\,\pdv{A}{\phi},\\
\Psi_{\tau_0}(\tau) &= \frac{1}{2\eta}\,\pdv{A}{\Omega},\\
\Psi_\Omega(\tau) &=-\frac{1}{2\eta}\,\pdv{A}{\tau_0}
+\frac{\Omega}{2\eta}\,\pdv{A}{\phi},\\
\Psi_\phi(\tau) &= \frac{1}{2}\,\pdv{A}{\eta}
-\frac{\Omega}{2\eta}\,\pdv{A}{\Omega}.
\end{aligned}
\end{equation}
These adjoint kernels are not identical to the tangent vectors of the ansatz manifold; rather, they are determined by the inverse restricted symplectic structure of the collective-coordinate basis. They specify how a field fluctuation is resolved into amplitude, frequency, timing, and phase forcing in the reduced coordinate space.

The ordering of the field-level noise enters precisely at the projection step.
Indeed, Eq.~\eqref{eq:proj_force_def} can be written as
\begin{equation}
F_i(t)
=
\frac{1}{2}\int_{-\infty}^{\infty}
\left[
\Psi_i^*(\tau)F(t,\tau)
+
\Psi_i(\tau)F^\dagger(t,\tau)
\right]\dd\tau .
\end{equation}
Thus, although the reduced forces $F_i(t)$ are real quadrature projections,
their covariance is still determined by the ordered correlations of the
underlying field noise. For vacuum input, Eqs.~\eqref{eq:noise_phys_vacuum}
and \eqref{eq:vacuum_ordering_phys} imply that only the contraction
$\langle F F^\dagger\rangle$ contributes. In contrast, for a commuting
classical circular complex white-noise source, the two contractions
$\langle F_{\rm cl}F_{\rm cl}^*\rangle$ and
$\langle F_{\rm cl}^*F_{\rm cl}\rangle$ are identical and both appear in the
covariance of the same real projection.
Consequently, a classical complex noise source with correlation strength
$B_{\rm cl}$ produces twice the projected real-quadrature diffusion that a
vacuum ordered correlator of the same numerical strength would produce.
Equivalently, to reproduce the vacuum-limited reduced diffusion using a
commuting classical Gaussian noise field, its field-level correlation strength
must be chosen as
\begin{equation}
B_{\rm cl}=\frac{B}{2}.
\end{equation}
This convention is the one used below when comparing the reduced vacuum
diffusion matrix with stochastic simulations based on classical complex
Gaussian noise.

For vacuum noise forcing, in the small-redshift regime, the vacuum diffusion matrix can therefore be written directly as [see Sec.~3 in Supplemental Documents]
\begin{equation}
\label{eq:Dmatrix_vacuum}
\mathbf{D}_{\rm(vac)}
\simeq
\mathrm{diag}\!\Bigl(D_{\eta\eta},\,D_{\Omega\Omega},\,D_{\tau_0\tau_0},\,D_{\phi\phi}\Bigr),
\end{equation}
with
\begin{equation}
\label{eq:Dii_vacuum}
\begin{aligned}
D_{\eta\eta}&=\frac{B\eta}{4},
&
D_{\Omega\Omega}&=\frac{B\eta}{12},
\\
D_{\tau_0\tau_0}&=\frac{B\pi^2}{48\eta^3},
&
D_{\phi\phi}&=\frac{B(1+\pi^2/12)}{12\eta}.
\end{aligned}
\end{equation}

This leading-order form shows that, in the chosen projected basis, vacuum fluctuations inject independent noise into the four collective-coordinate quadratures, while the correlations observed later in the spectra arise primarily from deterministic mixing through the drift matrix $\mathbf{M}$. More general noise models, or higher-order treatments beyond the small-redshift approximation, can be incorporated straightforwardly by retaining the off-diagonal entries of $\mathbf{D}$.


\subsection{PSD matrix and pathway decomposition}
\label{ssec:psd_matrix}

We now recast the linearized stochastic dynamics of Eq.~\eqref{eq:linear_system} in the frequency domain. This yields a closed spectral description of the collective-coordinate fluctuations.

For a stationary fluctuation vector, the frequency-domain second-order correlation is diagonal in frequency:
\begin{equation}
\label{eq:PSD_def_delta}
\bigl\langle
\Delta\mathbf{x}(\omega)\,
\Delta\mathbf{x}^{\dagger}(\omega')
\bigr\rangle
=
2\pi\,\delta_\mathrm{D}(\omega-\omega')\,\mathbf{S}_{\mathbf{x}}(\omega),
\end{equation}
which defines the two-sided PSD matrix \(\mathbf{S}_{\mathbf{x}}(\omega)\). The covariance matrix is recovered from it as
\begin{equation}
\label{eq:variance_from_psd}
\bigl\langle \Delta x_i\,\Delta x_j \bigr\rangle
=
\frac{1}{2\pi}
\int_{-\infty}^{\infty} S_{x_i x_j}(\omega)\, d\omega
=
\int_{-\infty}^{\infty} S_{x_i x_j}(f)\, df.
\end{equation}
Here \(x_i\) denotes the \(i\)-th reduced coordinate; for \(i=1,2,3,4\), these are \((\eta,\Omega,\tau_0,\phi)\). Throughout this work, all spectra are reported as two-sided PSDs; for \(\omega>0\), the corresponding one-sided PSD is \(S^{(1)}(\omega)=2S^{(2)}(\omega)\).

Fourier transforming Eq.~\eqref{eq:linear_system} gives
\begin{equation}
\label{eq:Delta_x}
\Delta\mathbf{x}(\omega)=\mathbf{H}(\omega)\,\mathbf{F}(\omega),
\end{equation}
with transfer matrix
\begin{equation}
\label{eq:H_def}
\mathbf{H}(\omega)=(-i\omega\mathbf{I}-\mathbf{M})^{-1},
\end{equation}
where \(\mathbf{I}\) is the identity matrix. Fourier transforming Eq.~\eqref{eq:D_def} and substituting the result into Eq.~\eqref{eq:PSD_def_delta} yields
\begin{equation}
\label{eq:PSD_matrix_general}
\mathbf{S}_{\mathbf{x}}(\omega)
=
\mathbf{H}(\omega)\,\mathbf{D}\,\mathbf{H}^{\dagger}(\omega).
\end{equation}
Equivalently, in component form,
\begin{equation}
\label{eq:PSD_matrix_general_component}
S_{x_i x_j}(\omega)
=
\sum_{k,l} H_{ik}(\omega)\,D_{kl}\,H_{jl}^*(\omega).
\end{equation}
Thus, the reduced spectra are determined jointly by the direct stochastic forcing statistics encoded in \(\mathbf{D}\) and the deterministic filtering and inter-coordinate coupling encoded in \(\mathbf{H}(\omega)\).

To convert coordinate spectra into pulse observables, we linearize the ansatz relations in Eq.~\eqref{eq_eta_to_pulse_paras}:
\[
\delta A_0=\sqrt{2}\,\delta\eta,\,\,\,\,
\delta I_0=4\eta\,\delta\eta,\,\,\,\,
\delta\tau_\mathrm{w}=-\eta^{-2}\delta\eta,\,\,\,\,
\delta E=4\,\delta\eta,
\]
where $I_0=|A_0|^2$.
Hence, for any observable $y(\eta)$ that depends only on $\eta$,
\[
S_{yy}(\omega)
=
\left(\frac{dy}{d\eta}\right)^2 S_{\eta\eta}(\omega).
\]
In particular,
\begin{subequations}
\label{eq:PSD_from_eta}
\begin{align}
\label{eq:PSD_A0_from_eta}
S_{A_0A_0}(\omega) &= 2\,S_{\eta\eta}(\omega), \\
\label{eq:PSD_I0_from_eta}
S_{I_0I_0}(\omega) &= 16\eta^2\,S_{\eta\eta}(\omega), \\
\label{eq:PSD_tauw_from_eta}
S_{\tau_\mathrm{w}\tau_\mathrm{w}}(\omega) &= \eta^{-4}\,S_{\eta\eta}(\omega), \\
\label{eq:PSD_E_from_eta}
S_{EE}(\omega) &= 16\,S_{\eta\eta}(\omega).
\end{align}
\end{subequations}
These relations show that amplitude-related observables inherit their full spectral structure directly from the single reduced spectrum $S_{\eta\eta}(\omega)$.

If a dimensional observable scales as
\[
\tilde x(\tilde t)=x_s\,x(t),
\qquad
\tilde t=t_\mathrm{s} t,
\]
then its PSD with respect to physical angular frequency $\tilde\omega$ is
\begin{equation}
\label{eq:psd_scaling_general}
S_{\tilde x\tilde x}(\tilde\omega)
=
x_s^2\,t_\mathrm{s}\,
S_{xx}\!\bigl(\omega=t_\mathrm{s}\tilde\omega\bigr),
\end{equation}
where $x_s$ is the corresponding normalization scale from Eq.~\eqref{eq:scales}. Thus, once the normalized PSD is known, its conversion to laboratory units is completely fixed by the same scaling dictionary that relates the reduced coordinates to the physical pulse observables.

As an example, the normalized peak-intensity relative-intensity-noise (RIN) is
\[
S_{\mathrm{RIN}}(\omega)
=
\frac{S_{I_0I_0}(\omega)}{I_0^2}
=
\frac{4}{\eta^2}S_{\eta\eta}(\omega),
\qquad
I_0=|A_0|^2=2\eta^2.
\]
Because the RIN is dimensionless, only the time rescaling enters when converting to physical units:
\begin{equation}
\label{eq:PSD_RIN_from_eta}
S_{\mathrm{RIN}}(\tilde\omega)
=
t_\mathrm{s}\,S_{\mathrm{RIN}}\!\bigl(t_\mathrm{s}\tilde\omega\bigr)
=
\frac{4t_\mathrm{s}}{\eta^2}\,
S_{\eta\eta}\!\bigl(t_\mathrm{s}\tilde\omega\bigr).
\end{equation}

Equations~\eqref{eq:PSD_matrix_general}--\eqref{eq:PSD_RIN_from_eta} provide the complete bridge from the linearized collective-coordinate Langevin system to measurable spectra. The following subsections use this framework to identify which pathways dominate the frequency, timing, amplitude, and phase noise over the stationary single-soliton regime.

\section{Pathway-resolved spectra, validation, and mapping to observables}
\label{sec:pathway_spectra}

\begin{figure*}[!t]
    \centering
    \includegraphics[width=\textwidth]{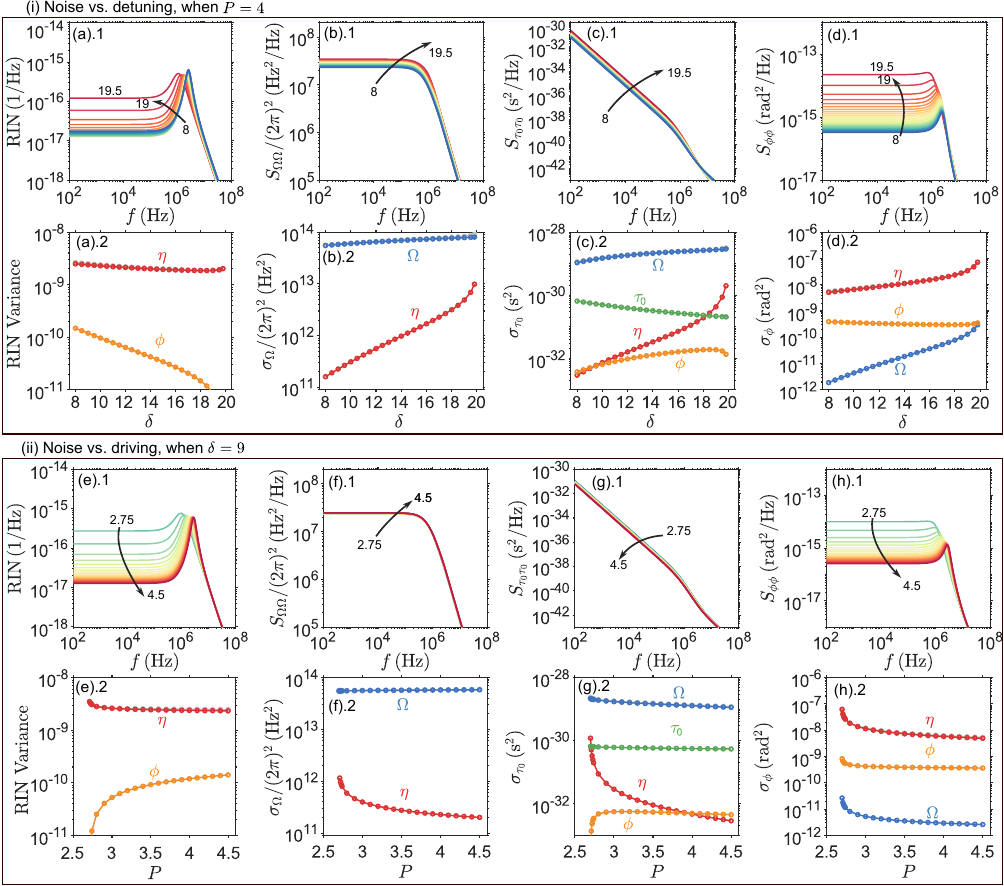}
    \caption{
    Detuning and driving dependence of the linearized spectra and the corresponding band-limited noise metrics. Physical units are derived from the silica-wedge scaling of Ref.~\cite{baoQuantumDiffusionMicrocavity2021} listed in Table~\ref{tab:scales}. In block (i), the detuning $\delta$ is swept from $\delta=8$ to $20$ at fixed $P=4$; in block (ii), the drive $P$ is swept from 2.7 to 4.5 at fixed $\delta=9$. In each block, panels .1 show analytic PSDs and panels .2 the corresponding integrals over $10^2~\mathrm{Hz}\le f\le 10^8~\mathrm{Hz}$. From left to right, the four columns correspond to the peak-intensity RIN, frequency-shift-noise PSD $S_{\Omega\Omega}$, timing-jitter PSD $S_{\tau_0\tau_0}$, and common-phase-noise PSD $S_{\phi\phi}$. The PSD curves are color-coded by the swept parameter, with arrows indicating the direction of increase. In panels .2, colored markers labeled by $\eta$, $\Omega$, $\tau_0$, and $\phi$ show the additive pathway contributions from the corresponding diffusion channels.
    }
    \label{fig:psd_sigma2_vs_parameters}
\end{figure*}

{We now use the linearized stochastic theory to determine how fluctuations are distributed across amplitude, frequency, timing, and phase, and to test how accurately the reduced description reproduces the corresponding spectra of the full LLE. The PSD-matrix formulation is useful not only because it predicts total spectra, but also because it resolves how each observable is assembled from direct injection and inter-coordinate conversion pathways. The logic of this section is therefore as follows. 
We first validate the analytic PSDs and their pathway decomposition against stochastic simulations of the reduced Langevin system and against spectra extracted from the full LLE. 
Having established that benchmark in the stationary fixed-point regime, we then use the reduced theory to track how the dominant pathways reorganize with detuning. 
To illustrate the platform-independent mapping of the normalized theory to physical units, we use two representative scalings from Table~\ref{tab:scales}. The validation example in Fig.~\ref{fig:PSD_decomposed} is shown using the SMF ring I scaling, whereas the parameter sweeps in Fig.~\ref{fig:psd_sigma2_vs_parameters} are shown using the silica-wedge scaling of Ref.~\cite{baoQuantumDiffusionMicrocavity2021}. The underlying dimensionless reduced model is the same in both cases; only the conversion to physical units differs.
Finally, we derive compact subsystem expressions that apply across resonator platforms and explain the two main classes of behavior seen in the sweep: broadband frequency--timing noise and resonance-like amplitude--phase noise.}

\subsection{Representative validation and mapping to LLE observables}
\label{ssec:example_psds}

\begin{figure*}[!t] 
\centering 
\includegraphics[width=\textwidth]{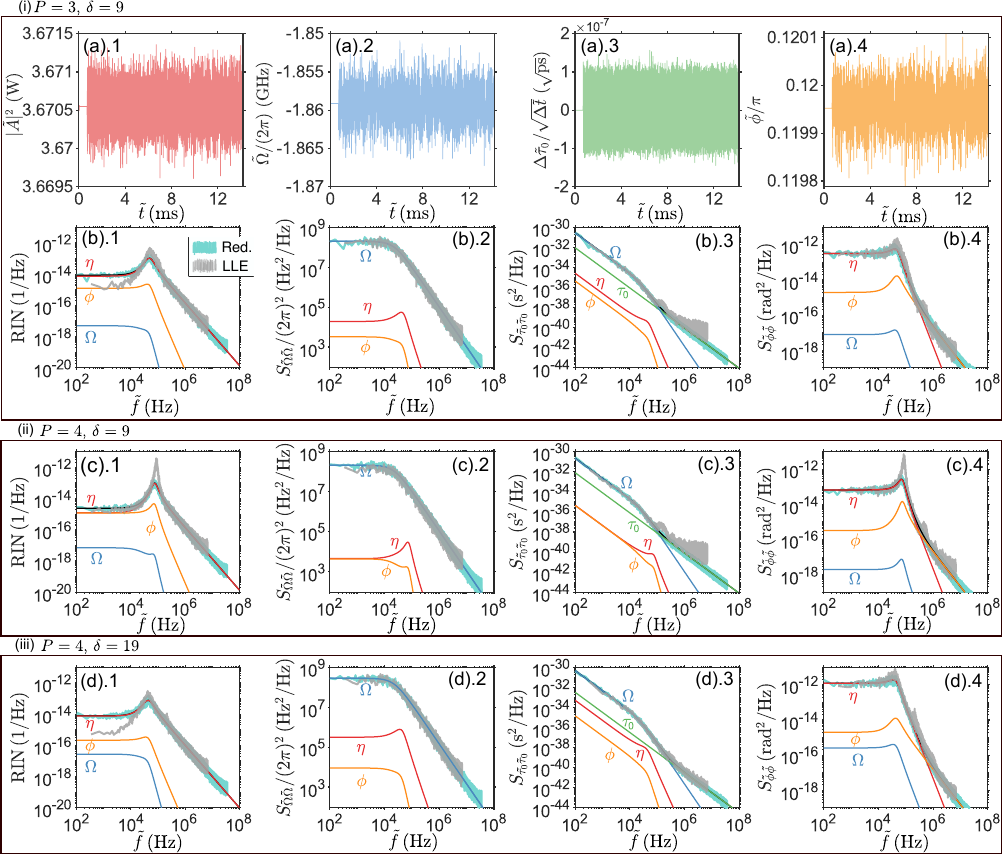} 
\caption{Representative comparisons of the linearized collective-coordinate Langevin theory against stochastic reduced-model and full-LLE simulations at three stationary operating points. Physical units are shown using the SMF ring scaling \cite{leoTemporalCavitySolitons2010} from Table~\ref{tab:scales}. Panels (a).1--4 show representative post-transient reduced-model trajectories at $(P,\delta)=(3,9)$ for the peak intensity $\max(|\tilde A|^2)$, frequency shift $\tilde\Omega/(2\pi)$, rescaled timing increment $(\tilde\tau_{0,n}-\tilde\tau_{0,n-1})/\sqrt{\Delta\tilde t}$, and common phase $\tilde\phi/\pi$. Panels (b).1--4 show the corresponding PSDs at $(P,\delta)=(3,9)$; panels (c).1--4 and (d).1--4 show the PSDs at $(P,\delta)=(4,9)$ and $(4,19)$, respectively. From left to right, the spectra are the peak-intensity RIN, $S_{\tilde\Omega\tilde\Omega}$, $S_{\tilde\tau_0\tilde\tau_0}$, and $S_{\tilde\phi\tilde\phi}$. Cyan shaded and light-gray curves show numerical PSDs from the reduced model (RM) and the full LLE, respectively; colored solid curves, labeled by $\eta$, $\Omega$, $\tau_0$, and $\phi$, show the individual pathway contributions from the corresponding diffusion channels; black curves show the total analytic prediction.
The origin of the different spectral shapes is analyzed in Secs.~\ref{ssec:PSD_tau0} and~\ref{ssec:eta_phi_resonance}.
}
\label{fig:PSD_decomposed} 
\end{figure*}

Having established the PSD matrix and its pathway decomposition in Sec.~\ref{ssec:psd_matrix}, we now test the theory at the level it is intended to describe. 
Figure~\ref{fig:PSD_decomposed} collects these comparisons at three representative stationary operating points: $(P,\delta)=(3,9)$, $(4,9)$, and $(4,19)$, corresponding to the points (i,ii,iii) in the phase diagram in Fig.~\ref{fig:detuning_vs_phase}(d). At any such point, the analytic theory makes two concrete predictions. First, Eq.~\eqref{eq:PSD_matrix_general} gives the total PSD of each collective coordinate. Second, that total PSD can be decomposed into additive contributions seeded by the individual diffusion channels $D_{\eta\eta}$, $D_{\Omega\Omega}$, $D_{\tau_0\tau_0}$, and $D_{\phi\phi}$. The figure is organized to test both predictions.

We begin with the reduced model at the representative operating point \((P,\delta)=(3,9)\), before comparing it with stochastic simulations of the
corresponding LLE. For this validation example, the normalized variables are converted to physical units using the SMF ring I scaling listed in Table~\ref{tab:scales}. Using Eqs.~\eqref{eq:param_odes} and \eqref{eq:G_components}, together with the vacuum diffusion matrix in Eqs.~\eqref{eq:Dmatrix_vacuum} and \eqref{eq:Dii_vacuum}, we initialize the collective coordinates at the stationary solution and first evolve the system with the Langevin forcing disabled. During this deterministic stage, up to \(\tilde t=50\,t_\mathrm{s}\simeq0.713~\mathrm{ms}\), the coordinates remain at the stationary fixed point, as shown in Fig.~\ref{fig:PSD_decomposed}(a).1--4. We then switch on the Langevin forcing, after which the peak intensity, frequency shift, rescaled timing increment, and common phase fluctuate around their stationary mean values.

For any coordinate $x(\tilde t)$, we estimate the PSD of the fluctuation
$\Delta x(\tilde t)=x(\tilde t)-\langle x\rangle$
over a finite analysis window by
\begin{equation}
S_{xx}(f)\equiv \frac{1}{T}\left|\int_{\tilde t_1}^{\tilde t_2}\Delta x(\tilde t)\,e^{i2\pi f \tilde t}\,d\tilde t\right|^2,
\qquad
T=\tilde t_2-\tilde t_1,
\label{eq_num_calcu}
\end{equation}
with simulation step $\Delta \tilde t = 10^{-3}t_\mathrm{s}$. To exclude the transient immediately after the noise is switched on, we analyzed traces between $\tilde t_1=100\,t_\mathrm{s}$ and $\tilde t_2=1000\,t_\mathrm{s}$, therefore the analyzed duration is
$T=900\,t_\mathrm{s}\simeq 12.834~\mathrm{ms}$, which sets the frequency resolution, or equivalently the lowest nonzero resolved Fourier frequency, to $1/T\approx 77.91~\mathrm{Hz}$. No window function was applied before the Fourier transform. The cyan curves in Fig.~\ref{fig:PSD_decomposed}(b)--(d) are averages over 10 independent realizations; for the reduced model, the full calculation requires only a few seconds on a standard computer.

The timing coordinate requires separate treatment because  $\tilde\tau_0(\tilde t)$ exhibits random-walk behavior, whereas its increments are stationary. Numerically, it is more robust to work with the rescaled discrete timing kicks
\begin{equation}
\kappa_n \equiv \frac{\tilde\tau_{0,n}-\tilde\tau_{0,n-1}}{\sqrt{\Delta\tilde t}},
\end{equation}
estimate their spectrum $S_{\kappa\kappa}$, and then reconstruct the timing-jitter PSD. In discrete time, the exact relation is
\begin{equation}
S_{\tilde\tau_0\tilde\tau_0}(\tilde\omega)
=
\frac{\Delta\tilde t\, S_{\kappa\kappa}(\tilde\omega)}
{|1-e^{-i\tilde\omega\Delta \tilde t}|^2}
=
\frac{\Delta\tilde t\, S_{\kappa\kappa}(\tilde\omega)}
{4\sin^2(\tilde\omega\Delta\tilde t/2)}\simeq \frac{S_{\kappa\kappa}(\tilde\omega)}{\tilde\omega^2}.
\label{eq_timing_cal_tricks_exact}
\end{equation}
For that reason, Fig.~\ref{fig:PSD_decomposed}(a).3 displays the rescaled timing-kick variable $\kappa_n$ rather than $\tilde\tau_0(\tilde t)$ itself. 

For connection to experimentally accessible RF observables, the timing-jitter spectrum can be converted into the phase-noise spectrum of the RF beat note (timing phase) via
\begin{equation}
S_{\tilde\psi_0\tilde\psi_0}(\tilde\omega)
=
\Omega_{\mathrm{r}}^2\,S_{\tilde\tau_0\tilde\tau_0}(\tilde\omega),
\label{eq_timing_jitter_to_timing_phase}
\end{equation}
assuming $\tilde\psi_0 = \Omega_{\mathrm{r}}\tilde\tau_0$. This timing-phase noise is distinct from the common-phase-noise spectrum $S_{\tilde\phi\tilde\phi}$ shown in Fig.~\ref{fig:PSD_decomposed}.

The PSD comparison at $(P,\delta)=(3,9)$ is shown in Fig.~\ref{fig:PSD_decomposed}(b).1--4. 
The solid colored curves are the pathway-resolved analytic contributions obtained from Eq.~\eqref{eq:PSD_matrix_general}, labeled by the seeding diffusion channels $\eta$, $\Omega$, $\tau_0$, and $\phi$, while the black curve gives the total prediction. 
The agreement with the reduced-model simulations is excellent across the full resolved frequency range. 
At this operating point, the pathway hierarchy is already clear: the RIN is dominated by direct amplitude injection, the frequency-noise spectrum by direct forcing of $\Omega$, the timing-jitter spectrum mainly by Gordon--Haus-type conversion $\Omega\to\tau_0$ at low offset, and the phase-noise spectrum predominantly by amplitude-to-phase conversion $\eta\to\phi$ rather than by direct phase diffusion. 
The spectra also display distinct shapes: the frequency-noise spectrum has a broadband Lorentzian-like response, the timing spectrum inherits a diffusive low-frequency scaling from the neutral coordinate $\tau_0$, whereas the RIN and phase spectra can develop resonance-like humps associated with the coupled amplitude--phase subsystem. 
These spectral morphologies and their analytic origins are discussed in detail in Secs.~\ref{ssec:PSD_tau0} and~\ref{ssec:eta_phi_resonance}. 
This comparison validates both ingredients of the linearized theory: the projection-derived diffusion matrix and the transfer-function structure encoded in $\mathbf{H}(\omega)$.

We next compare the same predictions with stochastic simulations of the full LLE [Eq.~\eqref{eq:LLE_phys} or Eq.~\eqref{eq:normalized}]. Because the LLE is an infinite-dimensional field equation, this comparison is meaningful only if the observables are extracted through field-level estimators that isolate the soliton contribution while minimizing sensitivity to the numerical grid and to the CW background. We therefore define the measured quantities through suitably windowed temporal or spectral moments of the field rather than through single grid-point values. 
In particular, the RIN is obtained from fluctuations of the field energy within a temporal window 10 times larger than the soliton duration, and is then converted to peak-intensity RIN using the reduced-model relations between energy and amplitude [Eqs.~\eqref{eq:PSD_E_from_eta} and \eqref{eq:PSD_RIN_from_eta}].
The timing jitter and frequency-shift fluctuations are extracted from power-weighted temporal and spectral centroids, respectively. For the phase, we use a weighted spectral-phase estimator designed to isolate the common soliton phase while suppressing contamination from timing jitter. 
Aside from a slight low-frequency discrepancy in the RIN, likely because the field-based estimator also includes fluctuations of the CW background, the agreement with the reduced model is very good for all other observables.
The explicit estimators and implementation details are given in Sec.~4 in the Supplemental Document.

The corresponding LLE spectra are the light-gray curves in Fig.~\ref{fig:PSD_decomposed}(b)--(d). They are averaged over 10 independent realizations obtained with $2^{13}$ fast-time grid points over a temporal window of $120\tau_\mathrm{s}$, a slow-time step $\Delta T=0.005t_\mathrm{s}\approx0.713\,ns$, and an analyzed duration $T=210t_\mathrm{s}$ for the Fourier transform. This analysis window sets the frequency resolution, or equivalently the lowest nonzero resolved Fourier frequency, to $1/T \approx 334~\mathrm{Hz}$. At $(P,\delta)=(3,9)$, the LLE spectra closely follow both the reduced-model simulations and the analytic predictions. This agreement is nontrivial: the LLE evolves the full field, requires a substantially smaller integration step, and retains fluctuation channels beyond the four-coordinate reduction. Its agreement with the reduced description therefore shows that, well inside the stationary-soliton regime, the dominant slow-time noise dynamics are already captured by the four collective coordinates and the linear conversion network that couples them. 
Thus, the reduced model is valuable not simply because it is faster, but because it reproduces the LLE spectral structure while revealing the noise-injection and conversion pathways that generate it.

The LLE timing-jitter spectra [gray curves] in Fig.~\ref{fig:PSD_decomposed}(b).3--(d).3 develop an approximately flat high-frequency floor. This is not a physical soliton-noise contribution, but a numerical artifact associated with the finite precision with which the pulse center can be localized on a discrete temporal grid. The physically relevant regime is $f<\mathrm{FSR}$ (0.539 MHz, see Tab.~\ref{tab:scales}), and this nearly flat part lies largely outside the range over which the mean-field description is expected to provide a reliable PSD prediction.

The other two operating points clarify the domain of validity of the reduction. When the operating point is moved to $(P,\delta)=(4,9)$ [Fig.~\ref{fig:PSD_decomposed}(c).1--4], closer to the Hopf boundary in Fig.~\ref{fig:detuning_vs_phase}, the CW background is stronger and its dynamical coupling to the soliton is no longer negligible. The full LLE then develops pronounced resonant features, most clearly in the RIN [Fig.~\ref{fig:PSD_decomposed}(c).1] and phase-noise spectra [Fig.~\ref{fig:PSD_decomposed}(c).4], while the frequency and timing spectra remain much better captured. These peaks indicate that an additional weakly damped degree of freedom becomes important near the onset of a secondary instability: the soliton couples more strongly to the background and to incipient breathing or internal-shape dynamics through energy exchange. 
Although such effects lie outside the present ansatz, the resulting noise-resonance peaks still anticipate the characteristic frequency scale of the breathing states that emerge beyond the stationary-soliton regime. Even in this regime, however, the reduced model continues to describe the frequency and timing sectors remarkably well. 
The amplitude and phase sectors exhibit a resonance-like spectral hump, captured by both the reduced model and the LLE. The corresponding peak frequency depends on the operating point $(P,\delta)$ and on the photon lifetime. This spectral peak should be distinguished from the damped eigenfrequency of the underlying amplitude--phase mode.
A more detailed analysis of the resonance-peak position is given in Sec.~\ref{ssec:eta_phi_resonance}.

When the operating point is moved away from the Hopf boundary to $(P,\delta)=(4,19)$ [Fig.~\ref{fig:PSD_decomposed}(d).1--4], the agreement for the RIN and phase-noise spectra recovers substantially. These trends in the evolution of the noise spectra will be quantified more systematically in the next subsection.

\subsection{Noise evolution vs detuning and driving}
\label{sec:noise_spectra_vs_detuning}

We now use the reduced theory to analyze how the noises evolve as the stationary single-soliton solution is varied through parameter space. In this subsection, physical units are mapped using the silica-wedge scaling of Ref.~\cite{baoQuantumDiffusionMicrocavity2021} listed in Table~\ref{tab:scales}. Figure~\ref{fig:psd_sigma2_vs_parameters} shows two complementary sweeps across the region identified by points (i)--(iii) in Fig.~\ref{fig:detuning_vs_phase}(d): varying detuning at fixed drive $P=4$ [block (i)], which moves along the stationary branch from (ii) toward (i), and varying drive at fixed detuning $\delta=9$ [block (ii)], which connects (iii) to (ii). In each block, panels .1 display the analytic PSDs, and panels .2 show the corresponding band-limited noise metrics obtained by integrating over the experimentally relevant offset-frequency range $10^2~\mathrm{Hz}\le f\le 10^8~\mathrm{Hz}$.

We first consider the detuning sweep at fixed $P=4$ [Fig.~\ref{fig:psd_sigma2_vs_parameters}(a--d)]. The clearest spectral change occurs in the amplitude--phase sector. In both the RIN and phase spectra [Fig.~\ref{fig:psd_sigma2_vs_parameters}(a).1 and (d).1], the finite-frequency hump remains present, but its contrast against the low-frequency background decreases as $\delta$ increases. Thus, increasing detuning does not remove noise from these observables; instead, it makes their spectra less visibly resonance-dominated. By contrast, the frequency-shift and timing spectra [Fig.~\ref{fig:psd_sigma2_vs_parameters}(b).1 and (c).1] retain their broadband character: $S_{\Omega\Omega}$ shows the familiar low-frequency plateau followed by a roll-off, while $S_{\tau_0\tau_0}$ keeps its diffusive $1/f^2$ form\cite{matskoTimingJitterMode2013,baoQuantumDiffusionMicrocavity2021}. In these two sectors, detuning mainly changes the overall level rather than the spectral shape.

The integrated metrics [Fig.~\ref{fig:psd_sigma2_vs_parameters}(a--d).2] show how this spectral evolution is assembled from the underlying pathways. The RIN remains dominated by direct amplitude injection through the $\eta$ channel, while the $\phi$-seeded correction decreases with detuning. The frequency-shift noise is dominated throughout by direct forcing of $\Omega$. The timing sector behaves differently: although the direct $\tau_0$ contribution decreases, the total timing noise increases because the dominant term is the converted $\Omega\to\tau_0$ pathway. Timing jitter therefore becomes increasingly conversion-dominated as $\delta$ grows. The phase-noise integral also rises strongly with detuning, and it does so mainly through the amplitude-seeded contribution; direct phase diffusion remains much smaller, and the $\Omega$-seeded term is only a secondary correction. Detuning therefore weakens the visible resonance in the amplitude--phase sector while simultaneously enhancing band-limited timing and phase fluctuations.

The drive sweep at fixed $\delta=9$ [Fig.~\ref{fig:psd_sigma2_vs_parameters}(e--h)] shows a different trend. As $P$ increases, the low-frequency parts of the RIN and phase spectra [Fig.~\ref{fig:psd_sigma2_vs_parameters}(e).1 and (h).1] are strongly suppressed, whereas the finite-frequency hump shifts to higher frequency. The main effect of drive is therefore to reduce low-offset amplitude and phase noise, not to reorganize the spectral topology. In contrast, the frequency-shift and timing spectra [Fig.~\ref{fig:psd_sigma2_vs_parameters}(f).1 and (g).1] vary only weakly across the sweep: $S_{\Omega\Omega}$ increases a little overall, and $S_{\tau_0\tau_0}$ keeps the same diffusive form with only a modest reduction in level.

This selectivity is made more explicit by the integrated pathway decomposition [Fig.~\ref{fig:psd_sigma2_vs_parameters}(e--h).2]. All contributions associated with $\eta$ decrease as the driving is increased. The RIN remains dominated by $\eta$ and decreases only moderately with increasing $P$ [Fig.~\ref{fig:psd_sigma2_vs_parameters}(e).2], despite the clear suppression of its low-frequency floor. The frequency-shift noise stays overwhelmingly dominated by $\Omega$ and is almost independent on the drive [Fig.~\ref{fig:psd_sigma2_vs_parameters}(f).2]. The timing integral likewise remains controlled mainly by the converted $\Omega\to\tau_0$ pathway and therefore also varies only weakly with $P$ [Fig.~\ref{fig:psd_sigma2_vs_parameters}(g).2]. By contrast, stronger driving reduces the phase noise [Fig.~\ref{fig:psd_sigma2_vs_parameters}(h).2], although it remains dominated by the amplitude-seeded channel.

These results also emphasize the importance of pathway-selective noise behavior for experimental design and applications. The detuning that minimizes one noise observable is not necessarily optimal for another. For example, the detuning that gives the lowest RIN is different from that which minimizes timing jitter [cf. Fig.~\ref{fig:psd_sigma2_vs_parameters}(a).2 and Fig.~\ref{fig:psd_sigma2_vs_parameters}(c).2]. Although one may identify an operating point that optimizes the overall noise performance, this optimum will generally differ from the detuning chosen when the goal is to minimize a specific quantity, such as RIN, timing jitter, or phase noise.

\subsection{Frequency and timing noise}
\label{ssec:PSD_tau0}

To gain further physical insight into the frequency and timing noise, we now analyze the corresponding reduced PSD structure. Rather than working with the full four-coordinate PSD matrix, it is more transparent to isolate the reduced subsystem that controls the coupled dynamics of $\Omega$ and $\tau_0$. Timing diffusion provides the clearest example: the neutral coordinate $\tau_0$ is driven both directly and indirectly through $\Omega$, while $\Omega$ itself can receive a Raman-mediated contribution from $\eta$. Over the parameter range considered here, the dominant structure is therefore captured by
\begin{subequations}
\begin{align}
\Delta\dot\eta &= -2\,\Delta\eta + F_\eta,
\label{eq:lin_eta}
\\
\Delta\dot\Omega &= m_{21}\,\Delta\eta -2\,\Delta\Omega + F_\Omega,
\label{eq:lin_Omega}
\\
\Delta\dot\tau_0 &= m_{32}\,\Delta\Omega + F_{\tau_0}.
\label{eq:lin_tau0}
\end{align}
\end{subequations}

Using Eq.~\eqref{eq:PSD_matrix_general}, the two-sided timing PSD separates into three additive contributions,
\begin{equation}
\label{eq:PSD_tau_total}
S_{\tau_0\tau_0}(\omega)
=
S_{\tau_0\tau_0}^{(\tau_0)}(\omega)
+
S_{\tau_0\tau_0}^{(\Omega)}(\omega)
+
S_{\tau_0\tau_0}^{(\eta)}(\omega),
\end{equation}
{corresponding to direct timing forcing, $\Omega\to\tau_0$ conversion, and the Raman-assisted chain $\eta\to\Omega\to\tau_0$. Explicitly,}
\begin{subequations}
\label{eq:PSD_tau_parts}
\begin{align}
S_{\tau_0\tau_0}^{(\tau_0)}(\omega)
&=
\frac{D_{\tau_0\tau_0}}{\omega^{2}}
=
\frac{\pi^{2}B}{48\eta^{3}}\,
\frac{1}{\omega^{2}},
\\
S_{\tau_0\tau_0}^{(\Omega)}(\omega)
&=
\frac{m_{32}^{2}D_{\Omega\Omega}}
          {\omega^{2}(\omega^{2}+4)}
=
\frac{B\eta}{12}\,
\frac{m_{32}^{2}}{\omega^{2}(\omega^{2}+4)},
\\
S_{\tau_0\tau_0}^{(\eta)}(\omega)
&=
\frac{m_{32}^{2}m_{21}^{2}D_{\eta\eta}}
         {\omega^{2}(\omega^{2}+4)^{2}}
=
\frac{B\eta}{4}\,
\frac{m_{32}^{2}m_{21}^{2}}{\omega^{2}(\omega^{2}+4)^{2}}.
\end{align}
\end{subequations}

All three terms inherit the diffusive $1/\omega^2$ factor from the neutral timing coordinate, while the Lorentzian factors reflect filtering by the damped Ornstein–Uhlenbeck processes. The direct timing term is controlled mainly by the soliton duration through $\eta^{-3}$, whereas the conversion terms depend on the operating point through $m_{32}$ and $m_{21}$. 

Applying the physical scaling of Eq.~\eqref{eq:psd_scaling_general} gives
\begin{subequations}
\begin{align}
S_{\tilde\tau_0\tilde\tau_0}^{(\tau_0)}(\tilde\omega)
&=
\frac{\pi^2\, B E_\mathrm{s}\, \alpha}{24\, \tilde E}\,
\frac{\tilde\tau_\mathrm{w}^2}{\tilde\omega^2},\label{eq:PSD_tau0_direct_phys}
\\
S_{\tilde\tau_0\tilde\tau_0}^{(\Omega)}(\tilde\omega)
&=
\frac{B\, \alpha^3\, \tau_\mathrm{s}^3\, m_{32}^2}
{96\, \tilde\tau_\mathrm{w}\, \tilde\omega^2\, (\alpha^2+\tilde\omega^2)},\label{eq:PSD_tau0_GH_jitter_phys}
\\
S_{\tilde\tau_0\tilde\tau_0}^{(\eta)}(\tilde\omega)
&=
\frac{32\, B\, \alpha^5\, \tau_\mathrm{s}^7\, \tau_\mathrm{R}^2\, m_{32}^2}
{225\, \tilde\tau_\mathrm{w}^7\, \tilde\omega^2\, (\alpha^2+\tilde\omega^2)^2}.\label{eq:PSD_tau0_Raman_phys}
\end{align}
\end{subequations}

When necessary, $E_{\rm s}$, $\tau_{\rm s}$, $\tilde E$, $\tilde A_0$, $\tilde \tau_\mathrm{w}$, and the scaling factors can be eliminated in favor of the relations given in Eqs.~\eqref{eq:duration_energy_relations} and \eqref{eq:scales}.
At the maximum-detuning point, where $m_{32}^2=4$, Eqs.~\eqref{eq:PSD_tau0_direct_phys} and \eqref{eq:PSD_tau0_GH_jitter_phys} reduce to the corresponding expressions reported in Ref.~\cite{baoQuantumDiffusionMicrocavity2021,matskoTimingJitterMode2013}.

Within the same subsystem, the frequency-shift PSD decomposes as
\begin{equation}
S_{\Omega\Omega}(\omega)
=
S_{\Omega\Omega}^{(\eta)}(\omega)+S_{\Omega\Omega}^{(\Omega)}(\omega),
\label{eq:PSD_Omega_decomp_onesided}
\end{equation}
{where}
\begin{subequations}
\label{eq:PSD_Omega_parts_vacuum_onesided}
\begin{align}
S_{\Omega\Omega}^{(\eta)}(\omega)
&=
\frac{m_{21}^{2}D_{\eta\eta}}{(\omega^{2}+4)^{2}}
=
\frac{1024\, B\, d_{R}^{2}\, \eta^{7}}
          {225\, (\omega^{2}+4)^{2}},
\\
S_{\Omega\Omega}^{(\Omega)}(\omega)
&=
\frac{D_{\Omega\Omega}}{\omega^{2}+4}
=
\frac{B\eta}{12(\omega^{2}+4)}.
\end{align}
\end{subequations}
{and the corresponding physical-unit expressions are}
\begin{subequations}
\label{eq:PSD_Omega_phys_onesided}
\begin{align}
S_{\tilde\Omega\tilde\Omega}^{(\eta)}(\tilde\omega)
&=
\frac{128\, B\, \alpha^3\, \tau_\mathrm{R}^2\, \tau_\mathrm{s}^3}
{225\, \tilde\tau_\mathrm{w}^7\, (\alpha^2+\tilde\omega^2)^2},
\\
S_{\tilde\Omega\tilde\Omega}^{(\Omega)}(\tilde\omega)
&=
\frac{B E_\mathrm{s}\, \alpha}
{6\, \tilde E\, \tilde\tau_\mathrm{w}^2\, (\alpha^2+\tilde\omega^2)}.
\end{align}
\end{subequations}

{The $(\eta,\Omega,\tau_0)$ reduction thus explains why the frequency spectra remain broadband and why the timing sector stays diffusive. 
Because this reduced subsystem suppresses the explicit role of $\phi$, however, it cannot explain the hump-like features of the RIN and phase spectra. Those require the coupled amplitude--phase block.}

\subsection{Amplitude--phase noise, resonance, and feedback}
\label{ssec:eta_phi_resonance}

{The detuning sweep showed that the RIN and phase spectra contain a finite-frequency hump whose contrast is strongest at low detuning and decreases as the stationary phase approaches zero. That feature is not generated by the diffusive timing sector, but by the coupled amplitude--phase dynamics. To expose this mechanism we retain the $(\eta,\Omega,\phi)$ block,}
\begin{equation}
\label{eq:lin_3param_eta_Omega_phi_general_alpha_noAj}
\Delta\dot\eta = -2\, \Delta\eta + m_{14}\, \Delta\phi + F_\eta, 
\end{equation}
\begin{equation}
\Delta\dot\Omega = m_{21}\, \Delta\eta -2\, \Delta\Omega + F_\Omega, 
\end{equation}
\begin{equation}
\Delta\dot\phi = m_{41}\, \Delta\eta + m_{42}\, \Delta\Omega -\frac{3}{2}\, \Delta\phi + F_\phi,
\end{equation}
{with diagonal diffusion matrix $\mathbf D=\mathrm{diag}(D_{\eta\eta}, D_{\Omega\Omega}, D_{\phi\phi})$.}

{The two-sided diagonal PSDs are}
\begin{subequations}
\label{eq:diag_PSDs_3param_general_alpha_noAj}
\begin{widetext}
\begin{align}
S_{\eta\eta}(\omega)
&=
\frac{
D_{\eta\eta}(\omega^2+4)\left(\omega^2+\frac{9}{4}\right)
+D_{\Omega\Omega}(m_{14}m_{42})^2
+D_{\phi\phi}\, m_{14}^2(\omega^2+4)
}{
|\Delta(\omega)|^2
},
\\
S_{\Omega\Omega}(\omega)
&=
\frac{
D_{\eta\eta}\, m_{21}^2{\left(\omega^2+\frac{9}{4}\right)}
+D_{\Omega\Omega}\, \Pi(\omega)
+D_{\phi\phi}(m_{14}m_{21})^2
}{
|\Delta(\omega)|^2
},
\\
S_{\phi\phi}(\omega)
&=
\frac{
D_{\eta\eta}\Bigl[(2m_{41}+m_{21}m_{42})^2+m_{41}^2\omega^2\Bigr]
+D_{\Omega\Omega}\, m_{42}^2(\omega^2+4)
+D_{\phi\phi}(\omega^2+4)^2
}{
|\Delta(\omega)|^2
}.
\end{align}
\end{widetext}
\end{subequations}
{where
\begin{equation}
\label{eq:B_modsq_noAj_simpler}
\Pi(\omega)=\Bigl|(2+i\omega)\Bigl(\frac{3}{2}+i\omega\Bigr)-m_{14}m_{41}\Bigr|^2.
\end{equation}}
{with common denominator}
\begin{equation}
\label{eq:Delta_modsq_noAj}
\begin{split}
|\Delta(\omega)|^2 
=& {(6-\frac{11}{2}\omega^2-2m_{14}m_{41}-C)^2} \\
+& \omega^2\bigl(10-\omega^2-m_{14}m_{41}\bigr)^2,
\end{split}
\end{equation}
where $C\equiv m_{14}m_{21}m_{42}$. These expressions separate two roles cleanly: the numerators determine how each Langevin channel seeds the observables, while the common denominator determines whether the coupled amplitude--phase block behaves as a broad overdamped response or develops an underdamped, resonance-like spectral response.

A particularly transparent limit is weak Raman conversion, $m_{21}\approx 0$, for which $\Omega$ becomes an almost independent Ornstein–Uhlenbeck process and $C\approx0$. In this regime,
\[
|\Delta(\omega)|^2=(\omega^2+4)\Pi(\omega).
\]
{In the additional limit of weak $\Omega\to\phi$ conversion, the amplitude and phase PSDs reduce to}
\begin{align}
S_{\eta\eta}(\omega)
&\simeq
\frac{
{\left(\omega^{2}+\frac{9}{4}\right)} D_{\eta\eta}
+m_{14}^{2}D_{\phi\phi}
}{
\Pi(\omega)
},
\\
S_{\phi\phi}(\omega)
&\simeq
\frac{
m_{41}^{2} D_{\eta\eta}
+(\omega^{2}+4) D_{\phi\phi}
}{
\Pi(\omega)
}.
\end{align}

{The corresponding physical-unit forms can be written as}
\begin{equation}
S_\mathrm{RIN}(\tilde\omega)
=
\frac{2B\alpha\frac{\tilde\tau_\mathrm{w}}{\tau_\mathrm{s}}}{\mathcal D_\alpha(\tilde\omega)}
\left[
{\left(\frac{9}{4}\alpha^2+4\tilde\omega^2\right)}
+
\frac{\alpha^{2}C_2m_{14}^{2}\tilde\tau_\mathrm{w}^2}{3\tau_\mathrm{s}^2}
\right],
\label{eq:RIN_phys_tauw_factored}
\end{equation}
\begin{equation}
S_{\tilde\phi\tilde\phi}(\tilde\omega)
\simeq
\frac{2B\alpha\frac{\tilde\tau_\mathrm{w}}{\tau_\mathrm{s}}}{\mathcal D_\alpha(\tilde\omega)}
\left[
\frac{\alpha^{2}m_{41}^{2}\tau_\mathrm{s}^2}{4\tilde\tau_\mathrm{w}^2}
+
\frac{C_2(\alpha^{2}+\tilde\omega^{2})}{3}
\right],
\end{equation}
{where}
\(
\mathcal D_\alpha(\tilde\omega)=
\Bigl((3-m_{14}m_{41})\alpha^2-4\tilde\omega^2\Bigr)^2
+49\alpha^2\tilde\omega^2
\) and
\(
C_2=1+\frac{\pi^2}{12}.
\)

\begin{figure}[t]
    \centering
    \includegraphics[scale=1]{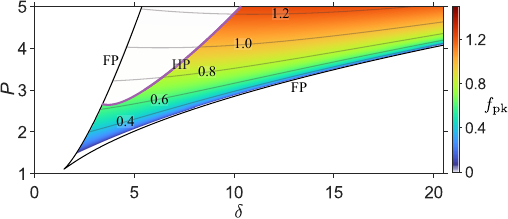}
    \caption{Evolution of the intensity- and phase-noise peak frequency $f_{\rm pk}$ in the $(P,\delta)$ plane.}
    \label{fig:f_peak_m14m41_PSD}
\end{figure}

\begin{figure*}[t]
    \centering
    \includegraphics[scale=1]{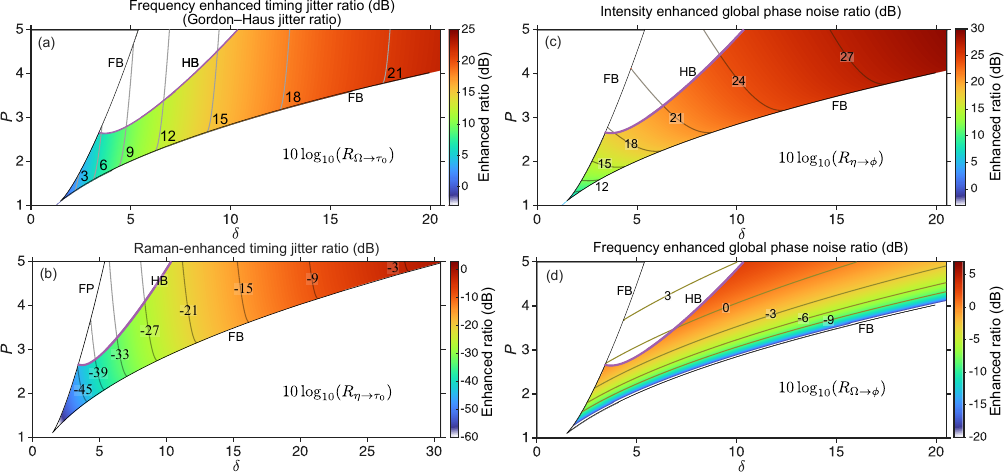}
    \caption{
    Low-frequency noise-enhancement ratios across the single-soliton existence region in the $(P,\delta)$ plane: (a) frequency-enhanced timing-jitter ratio ($\Omega\!\to\!\tau_0$) relative to direct timing diffusion; (b) Raman-assisted timing-jitter ratio ($\eta\!\to\!\Omega\!\to\!\tau_0$) relative to direct timing diffusion; (c) intensity-enhanced global-phase-noise ratio ($\eta\!\to\!\phi$) relative to direct phase diffusion; and (d) frequency-enhanced global-phase-noise ratio ($\Omega\!\to\!\phi$) relative to direct phase diffusion. Solid black curves indicate the soliton existence boundaries; the purple curve denotes the Hopf boundary. Panel (b) is evaluated for $\tau_\mathrm{s}=200~\mathrm{fs}$, which fixes $d_\mathrm{R}$ via Eq.~\eqref{eq_normalized_parameters}.
    }
    \label{fig:enhenced_ratio}
\end{figure*}

\emph{Resonance interpretation.} 
A finite-frequency hump is associated with the underdamped amplitude--phase subsystem. When the closed loop \(
\Delta\phi \xrightarrow{m_{14}} \Delta\eta \xrightarrow{m_{41}} \Delta\phi
\) supports a damped oscillatory eigenmode, the driven response can develop a nonzero-frequency PSD maximum. 
Detuning suppresses the feedback contrast of the resonance even though the integrated phase noise can continue to grow through amplitude-to-phase routing, which is precisely the trend identified in Fig.~\ref{fig:psd_sigma2_vs_parameters}(a,d). Since the numerator varies slowly across the resonance width, the spectral peak frequency is estimated by minimizing the reduced amplitude--phase denominator $\Pi(\omega)$, which gives
\begin{equation}
\label{eq:omega_min_noAj}
\omega_{\min}^2
\approx
{- m_{14}m_{41}-\frac{25}{8}},
\end{equation}
{and therefore}
\begin{equation}
\label{eq:fpk_dimless}
f_{\rm pk}
\approx\frac{1}{2\pi}\sqrt{{- m_{14}m_{41}-\frac{25}{8}}}.
\end{equation}
For the reduced $(\eta,\phi)$ subsystem, the damped eigenfrequency is instead determined by the imaginary part of the corresponding complex eigenvalues and satisfies
\(
\omega_{\rm eig}^2 = -m_{14}m_{41}-\frac{1}{16}.
\)
This frequency characterizes the free decaying transient, whereas Eq.~\eqref{eq:omega_min_noAj} estimates the maximum of the noise-driven real-frequency response. Thus, the existence of a damped oscillatory eigenmode is not by itself sufficient to guarantee a nonzero-frequency PSD maximum.

{Using $\tilde\omega=\omega f_\mathrm{s}$ with $f_\mathrm{s}=\alpha/2$, the corresponding physical peak frequency is}
\begin{equation}
\tilde f_{\rm pk}= f_{\rm pk}\frac{\alpha}{2}.
\end{equation}

Figure~\ref{fig:f_peak_m14m41_PSD} shows how this peak frequency evolves across the stationary-soliton region in the $(P,\delta)$ plane. The peak frequency is smallest near the lower-left edge of the existence region and increases steadily toward larger $P$ and, more weakly, toward larger $\delta$. 

\subsection{Low-frequency pathway-enhancement ratios}
\label{sssec:noise_enhance_ratio}

The preceding analysis established both the spectral structure of the noise and the detuning-driven reordering of its dominant pathways. To isolate this pathway hierarchy in the low-frequency limit, it is useful to introduce a quantity that removes the detailed spectral lineshape and compares only the strengths of competing injection channels. We therefore define the pathway-enhancement ratios
\begin{equation}
R_{j\to i}(\omega)\equiv
\frac{S_{x_ix_i}^{(j)}(\omega)}{S_{x_ix_i}^{(i)}(\omega)},
\qquad
S_{x_ix_i}^{(j)}(\omega)=|H_{ij}(\omega)|^2 D_{jj},
\end{equation}
where $S_{x_ix_i}^{(j)}$ denotes the contribution to the PSD of observable $x_i$ arising from diffusion injected in coordinate $x_j$. Thus, $R_{j\to i}>1$ means that the indirect pathway seeded in $x_j$ contributes more noise to $x_i$ than the direct diffusion of $x_i$ itself, whereas $R_{j\to i}<1$ means that the direct channel remains dominant. We focus on timing jitter ($i=\tau_0$) and collective phase noise ($i=\phi$), for which indirect conversion is most important in the present system.

For timing jitter, the dominant indirect contribution is the Gordon--Haus-type conversion $\Omega\to\tau_0$. Using the low-frequency asymptote of the transfer matrix gives
\begin{equation}
\label{eq:ratio_GH_direct}
R_{\Omega\to\tau_0}
\equiv
\frac{S_{\tau_0\tau_0}^{(\Omega)}(\omega)}{S_{\tau_0\tau_0}^{(\tau_0)}(\omega)}
\;\xrightarrow[\omega\to 0]{}\;
\frac{m_{32}^{2}D_{\Omega\Omega}/4}{D_{\tau_0\tau_0}}
=
\frac{\eta^{4}m_{32}^{2}}{\pi^{2}}.
\end{equation}
This result shows that the low-frequency timing-jitter enhancement scales as $\eta^{4}$ and is further controlled by the operating-point-dependent $m_{32}^{2}$.

A second indirect route is the Raman-assisted cascade $\eta\to\Omega\to\tau_0$, mediated by the coupling coefficient $m_{21}$:
\begin{equation}
\label{eq:R_eta_to_tau_direct}
R_{\eta\to\tau_0}
\equiv
\frac{S_{\tau_0\tau_0}^{(\eta)}(\omega)}{S_{\tau_0\tau_0}^{(\tau_0)}(\omega)}
\;\xrightarrow[\omega\to 0]{}\;
\frac{m_{32}^2 m_{21}^2 D_{\eta\eta}}{16D_{\tau_0\tau_0}}
=
\frac{3\eta^4m_{32}^2 m_{21}^2}{4\pi^2}.
\end{equation}

The corresponding maps of $10\log_{10}(R)$ are shown in Fig.~\ref{fig:enhenced_ratio}(a,b). Panel~(a) shows that the Gordon--Haus ratio is larger than 1 throughout the soliton existence region and increases strongly with both detuning and pump power. Timing jitter is therefore conversion-dominated over essentially the entire stationary-soliton regime: low-frequency timing noise is injected more efficiently through frequency fluctuations than through direct timing diffusion, and this dominance becomes progressively stronger at larger $\delta$ and $P$. Panel~(b) shows that the Raman-assisted ratio follows the same overall trend but $R$ remains smaller than 1 (i.e. $10\log(R)<0$) in the present parameter range for $\tau_\mathrm{s}=200~\mathrm{fs}$. It then becomes steadily less negligible toward larger detuning and pump power through the factor $m_{21}^{2}$ that feeds amplitude noise into timing jitter.

For the collective phase coordinate, the corresponding low-frequency enhancement ratios for $\eta$-seeded and $\Omega$-seeded phase noise are
\begin{align}
R_{\eta\to\phi}(\omega)\label{eq_R_eta_to_phi}
&\equiv
\frac{S_{\phi\phi}^{(\eta)}(\omega)}{S_{\phi\phi}^{(\phi)}(\omega)}
\;\xrightarrow[\omega\to 0]{}\;
\frac{D_{\eta\eta}}{D_{\phi\phi}}\,
\frac{(2m_{41}+m_{21}m_{42})^2}{16},
\\
R_{\Omega\to\phi}(\omega)
&\equiv
\frac{S_{\phi\phi}^{(\Omega)}(\omega)}{S_{\phi\phi}^{(\phi)}(\omega)}
\;\xrightarrow[\omega\to 0]{}\;
\frac{D_{\Omega\Omega}}{D_{\phi\phi}}
\frac{m_{42}^2}{4}.
\end{align}
The ratio $R_{\eta\to\phi}$ is dominated by the direct $m_{41}^{2}$ pathway and scales as $\eta^{4}$, while the mixed term $m_{21}m_{42}$ captures the Raman-mediated contribution to phase fluctuations.
Their distributions are shown in Fig.~\ref{fig:enhenced_ratio}(c,d). The amplitude-seeded ratio $R_{\eta\to\phi}\gg 1$ across almost the entire single-soliton region, typically by more than an order of magnitude in linear scale, showing that low-frequency global phase noise is governed primarily by amplitude-to-phase conversion. This enhancement becomes stronger as both detuning and pump power increase. By contrast, the frequency-seeded phase ratio $R_{\Omega\to\phi}$ is much weaker and more nonuniform: it is positive only over the lower-detuning, higher-power side of the existence region, but with $R<1$ over a broad high-detuning portion of the map. Frequency-to-phase conversion is therefore generally a secondary pathway, although it can become locally important close to the oscillatory instability.
This hierarchy is useful when considering noise-mitigation strategies. 
In particular, reducing timing jitter requires suppressing frequency-shift fluctuations, whereas reducing collective phase noise requires suppressing intensity fluctuations.

\subsection{From collective coordinates to optical-tooth, RF, and common-mode phase observables}
\label{ssec:global_comb_ceo}

The reduced description is formulated in terms of the collective coordinates
$\eta$, $\Omega$, $\tau_0$, and $\phi$, whereas experiments probe the system
through observables such as RIN, optical comb-tooth phase, photodetected RF
phase, and common-mode optical phase
\cite{Kim2016reviewModelockedLasers,DelHaye2016PhaseCoherent,Fortier2019_20YearsDevelopments}.
This subsection establishes the leading-order correspondence between the
collective coordinates and these experimental observables. The relations derived
below specify how fluctuations of $\eta$, $\Omega$, $\tau_0$, and $\phi$ appear
in measured quantities, regardless of the noise source that generates those
fluctuations. Small changes of the physical repetition rate caused, for example, by Raman-induced carrier-frequency shifts are not treated as a separate observable channel here; their leading effect is represented through the reduced timing dynamics, in particular through frequency-to-timing coupling. Amplitude fluctuations associated with $\eta$ map directly onto observables such
as RIN, pulse-duration noise, and energy noise through Eqs.~\eqref{eq:PSD_from_eta}; the discussion below therefore focuses on phase observables.

We expand the cavity field in longitudinal modes as
\begin{equation}
A(t,\tau)=\sum_{\mu\in\mathbb{Z}} a_\mu(t)\,e^{i\mu\Omega_\mathrm{r}\tau},
\qquad
\theta_\mu(t)\equiv \arg a_\mu(t),
\label{eq_modalexp}
\end{equation}
where $a_\mu(t)$ is the complex amplitude of the $\mu$th cavity mode,
equivalently the $\mu$th optical comb tooth in the pump-referenced frame. Under
the localized-pulse approximation, substituting the ansatz of
Eq.~\eqref{eq:nor_ansatz} into the Fourier coefficients and shifting the
integration variable by $\tau_0(t)$ yields
\begin{equation}
a_\mu(t)\approx
e^{i\phi(t)}\,e^{-i\mu\Omega_\mathrm{r}\tau_0(t)}
\frac{\Omega_\mathrm{r}}{\sqrt{2}}\,
\sech\!\left[
\frac{\pi(\mu\Omega_\mathrm{r}+\Omega)}{2\eta(t)}
\right].
\label{eq:amu_from_ansatz_main}
\end{equation}
For the present chirp-free ansatz, the final factor is real, so it does not contribute to the modal phase. Hence we have
\begin{equation}
\Delta\theta_\mu(t)\approx
\Delta\phi(t)-\mu\Omega_\mathrm{r}\,\Delta\tau_0(t).
\label{eq:theta_mu_affine_main}
\end{equation}
At the same leading order, $\Omega$ shifts the spectral envelope in
Eq.~\eqref{eq:amu_from_ansatz_main} but does not modify the affine phase
structure itself.

Direct photodetection measures the intensity profile rather than the optical
field phase. The periodic intensity can be expanded as
\begin{equation}
I(t,\tau)=|A(t,\tau)|^2
=\sum_{n\in\mathbb Z} b_n(t)e^{in\Omega_\mathrm{r}\tau},
\label{eq:intensity_RF_expansion_main}
\end{equation}
with
\begin{equation}
b_n(t)=\sum_\mu a_{\mu+n}(t)a_\mu^*(t)\approx
e^{-i n\Omega_\mathrm{r}\tau_0(t)}
\sum_\mu |a_{\mu+n}(t)|\,|a_\mu(t)|.
\label{eq:bn_exact_main}
\end{equation}
Here the mode-independent optical phase cancels in each beat note
$a_{\mu+n}a_\mu^*$, leaving only the timing-dependent phase factor.
Experimentally, the phase of the photodetected RF harmonic is measured relative
to a microwave reference, so that the measured RF phase noise is directly the
timing-phase noise. Accordingly,
\begin{equation}
\Delta\Phi_n^{(\mathrm{lab})}(t)\approx
-n\Omega_\mathrm{r}\,\Delta\tau_0(t),
\quad
S_{\Phi_n\Phi_n}^{(\mathrm{lab})}(\omega)\simeq
n^2\Omega_\mathrm{r}^2\,S_{\tau_0\tau_0}(\omega).
\label{eq:RF_phase_PSD_main}
\end{equation}

To access the common-mode optical phase without relying on the pumped tooth
itself, one may form a phase combination that removes the component linear in
the pump-referenced mode index $\mu$. The simplest choice is a symmetric pair
of comb teeth at relative indices $+\mu$ and $-\mu$. Since the timing-induced
phase is odd in $\mu$, whereas the common phase is even, the symmetric average
cancels the leading timing contribution:
\begin{equation}
\Delta\theta^{(\mathrm{sym})}(t)
\equiv
\frac{1}{2}
\left[
\Delta\theta_{+\mu}(t)
+
\Delta\theta_{-\mu}(t)
\right]
\approx
\Delta\phi(t).
\label{eq:sym_phase_def_main}
\end{equation}
The symmetry is taken with respect to the pumped mode $\mu=0$, not with
respect to the possibly redshifted center of the soliton spectrum. A Raman
redshift can displace the spectral envelope, but the phase origin of the
affine relation in Eq.~\eqref{eq:theta_mu_affine_main} remains the pumped
mode. More generally, two non-symmetric teeth $\mu_1$ and $\mu_2$ may also be
used if one forms the intercept at $\mu=0$,
\begin{equation}
\Delta\theta^{(0)}(t)
=
\frac{
\mu_2\Delta\theta_{\mu_1}(t)
-
\mu_1\Delta\theta_{\mu_2}(t)
}{
\mu_2-\mu_1
}
\approx
\Delta\phi(t).
\label{eq:common_phase_intercept_main}
\end{equation}
This generalized construction cancels the leading timing term, but can amplify
measurement noise when the two teeth are not well separated around $\mu=0$.
The resulting PSD therefore yields the collective common-phase spectrum
$S_{\phi\phi}(\omega)$, up to residual non-affine spectral-phase corrections.

\section{Conclusions}\label{sec_conclusions}

We have developed a pathway-resolved stochastic collective-coordinate theory for fluctuations of stationary driven--dissipative solitons in the generalized Lugiato--Lefever equation with Raman response. By projecting field-level vacuum fluctuations onto four soliton coordinates---amplitude $\eta$, frequency shift $\Omega$, timing $\tau_0$, and global phase $\phi$---we obtained both a nonlinear reduced Langevin model and, after linearization around a stable fixed point, an analytic PSD matrix. The central advance is not only the dimensional reduction itself, but the explicit separation between direct stochastic injection, encoded in the projected diffusion matrix, and deterministic inter-coordinate transfer, encoded in the Jacobian. This makes it possible to identify how each observable spectrum is assembled from distinct internal fluctuation pathways.

Within the stable stationary single-soliton regime, the theory yields a clear hierarchy of fluctuation transfer. Timing jitter remains diffusive because $\tau_0$ is neutral, but over much of the existence region its low-frequency spectrum is governed primarily by frequency-to-timing conversion $(\Omega\!\to\!\tau_0)$ rather than by direct timing diffusion. Frequency-noise spectra are dominated mainly by direct forcing of $\Omega$, with Raman-enabled $(\eta\!\to\!\Omega)$ transfer providing a secondary correction. By contrast, global phase noise is often governed less by direct phase diffusion than by amplitude-to-phase conversion $(\eta\!\to\!\phi)$. The resonance hump in the intensity and phase spectra is traced to the driven response of the coupled amplitude--phase subsystem, while Raman response opens additional cascaded routes, especially $(\eta\!\to\!\Omega)$ and $(\Omega\!\to\!\phi)$, that may reshape the timing and phase spectra in a parameter-dependent way. The parameter sweeps further show that detuning and drive reorganize these pathways differently: increasing detuning weakens the visible resonance hump contrast while strengthening band-limited timing and phase noise through conversion channels, whereas increasing drive shifts and sharpens the amplitude--phase resonance while reducing the overall amplitude, timing, and global phase noise.

Comparison with stochastic simulations of both the reduced model and the generalized LLE shows that this four-coordinate description captures the dominant slow-time fluctuation physics throughout most of the stable stationary regime. The agreement is especially strong for the frequency and timing sectors. Near the low-detuning Hopf boundary, the comparison also clarifies the limit of the present reduction: resonance features associated with the nearby breathing instability are already reflected in the reduced spectra, but quantitatively complete agreement in the amplitude and phase sectors ultimately requires additional degrees of freedom associated with background and shape dynamics.

The comparison between the fiber-ring and microresonator examples clarifies what is universal and what is platform-specific. For a given normalized operating point, the fluctuation-routing structure is set by the same reduced stochastic model, whereas the mapping to laboratory units depends on the platform-specific normalization scales. In particular, the cavity-loss scale sets the physical offset-frequency axis and relaxation bandwidth, while $\tau_\mathrm{s}$, $A_\mathrm{s}$, $E_\mathrm{s}$, and the normalized Raman parameter $d_\mathrm{R}=\tau_\mathrm{R}/\tau_\mathrm{s}$ determine the conversion of pulse parameters, absolute noise magnitudes, and Raman-mediated couplings. In this sense, fiber resonators and microresonators realize the same fluctuation-routing framework under different physical scalings rather than different underlying mechanisms. Through the mapping from collective coordinates to comb observables, this also connects the internal routing picture directly to measurable RIN, RF timing-phase noise, and common mode phase noise.

More broadly, the present results suggest that fluctuations of a dissipative localized state can be organized systematically as noise transfer on a reduced manifold of collective coordinates, provided that the dynamics remain sufficiently close to a stationary attractor. The most immediate extensions are to include technical and colored noise sources, enlarge the reduced state space to capture chirp and background modes, and generalize the framework to multisoliton and breathing states. More generally, the combination of stochastic projection and pathway-resolved spectra provides a practical route from microscopic forcing to experimentally accessible noise observables in driven--dissipative soliton systems.

\section{Acknowledgment}

{This work was supported by the Marie Sk{\l}odowska-Curie Actions (101149506, 101150387), the ERC Consolidator Grant (101125625), the F.R.S.-FNRS project EQP (40021523), and the Excellence of Science (EOS, 40007560) programme of FWO and F.R.S.-FNRS.}

\section*{References}
\bibliography{reference_organized3}

\end{document}